\documentclass[12pt]{article}
\usepackage{jheppub}

\usepackage{amsmath,bbm,array,amsfonts,graphicx,wrapfig,lscape,float,slashbox,multirow,longtable,rotating,subfigure,epstopdf}

\newcommand{\be}{\begin{equation}}
\newcommand{\ee}{\end{equation}}
\newcommand{\beq}{\begin{equation}}
\newcommand{\beql}[1]{\begin{equation}\label{#1}}
\newcommand{\eeq}{\end{equation}}
\newcommand{\ba}{\begin{array}}
\newcommand{\ea}{\end{array}}
\newcommand{\bea}{\begin{eqnarray}}
\newcommand{\beal}[1]{\begin{eqnarray}\label{#1}}
\newcommand{\eea}{\end{eqnarray}}
\newcommand{\ben}{\begin{enumerate}}
\newcommand{\een}{\end{enumerate}}
\newcommand{\bean}{\begin{eqnarray*}}
\newcommand{\eean}{\end{eqnarray*}}

\newcommand{\nn}{\nonumber}

\newcommand{\btab}[1]{\begin{tabular}{#1}}
\newcommand{\etab}{\end{tabular}}
\newcommand{\diag}{\mbox{diag}}

\newcommand{\comment}[1]{}

\newcommand{\qed}{\nobreak \ifvmode \relax \else
      \ifdim\lastskip<1.5em \hskip-\lastskip
      \hskip1.5em plus0em minus0.5em \fi \nobreak
      \vrule height0.75em width0.5em depth0.25em\fi}

\def\beqa{\begin{eqnarray}}
\def\eeqa{\end{eqnarray}}

\newcolumntype{C}[1]{>{\centering\arraybackslash}m{#1}}

\def\II{\relax{\rm I\kern-.18em I}}

\def\makeatletter{\catcode`\@=11}
\makeatletter
\def\mathbox#1{\hbox{$\m@th#1$}}%
\def\math@ccstyles#1#2#3#4#5#6#7{{\leavevmode
     \setbox0\mathbox{#6#7}%
     \setbox2\mathbox{#4#5}%
     \dimen@ #3%
     \baselineskip\z@\lineskiplimit#1\lineskip\z@
     \vbox{\ialign{##\crcr
            \hfil \kern #2\box2 \hfil\crcr
            \noalign{\kern\dimen@}%
            \hfil\box0\hfil\crcr}}}}
\def\mathaccstyles{\math@ccstyles\maxdimen}
\def\maththroughstyles{\math@ccstyles{-\maxdimen}}
\def\unity%
{\maththroughstyles{.45\ht0}\z@\displaystyle {\mathchar"006C}\displaystyle 1}

\title{Orientifolds of Warped Throats from \\ Toric Calabi-Yau Singularities}

\author[a,b]{Ander Retolaza}
\author[a]{, Angel Uranga}

\affiliation[a]{Instituto de F\'isica Te\'orica UAM-CSIC \\
C/ Nicol\'as Cabrera 13-15, Campus de Cantoblanco,  28049 Madrid, Spain}
\affiliation[b]{ Departamento de F\'isica Te\'orica, Universidad Aut\'onoma de Madrid,\\ Campus de Cantoblanco, 28049 Madrid, Spain }
\emailAdd{ander.retolaza@uam.es, angel.uranga@uam.es}

\abstract{We study the complex deformations of orientifolds of D3-branes at toric CY singularities, using their description in terms of dimer diagrams. We describe orientifold quotients that have fixed lines or fixed points in the dimer, and characterize the possibilities to deform them in terms of the behaviour of zig-zag paths under the orientifold symmetry. The resulting models are holographic duals to warped throats with orientifold planes. Our systematic construction provides a general class of configurations which includes models recently appeared in the context of de Sitter uplift by nilpotent goldstino or dynamical supersymmetry breaking.
}

\preprint{
\begin{flushright}IFT-UAM/CSIC-16-041 \\ FTUAM-16-15\end{flushright} \vspace{-0.9cm}
}

\begin{document}

\maketitle


\section{Introduction and main results} \label{sec:intro}

Warped throats are an interesting ingredient in string compactifications, as they provide a mechanism to generate hierarchies in sectors localized in the internal dimensions.  For instance,  contributions to the vacuum energy proposed to uplift to de Sitter vacua \cite{Kachru:2003aw}, brane inflation \cite{Kachru:2003sx} or certain axion monodromy inflation models \cite{Franco:2014hsa,Retolaza:2015sta}, particle physics models \cite{Cascales:2003wn,Cascales:2005rj,Franco:2008jc}, etc. used these type of geometries.  Although the prototypical example of a warped throat is based on the deformed conifold and its holographic dual RG flow in terms of a duality cascade \cite{Klebanov:2000hb}, it is easy to generalize these ideas to produce many other throats based on complex deformations of other toric CY singularities, which also admit   duality cascades  on the holographic dual description \cite{Franco:2004jz,Franco:2005fd}.

In compact examples, cancellation of the RR charges carried by the fluxes supporting the throat forces the introduction of orientifold planes \cite{Giddings:2001yu}. These are usually located away from the throat, and therefore their presence is irrelevant to the infrared physics down the throat. However, some recent applications exploit the presence of orientifold planes at the infrared tip of the throat, i.e. they involve orientifolded warped throats. This has appeared in the gauge theory description of certain D-brane instantons \cite{Aharony:2007pr,Amariti:2008xu,Franco:2015kfa}, and de Sitter uplifts using string embeddings of the nilpotent goldstino \cite{Kallosh:2015nia} and DSB sectors \cite{Retolaza:2015nvh}. 

The construction of orientifolds of D3-branes at toric CY singularities was systematized in \cite{Franco:2007ii}, based on dimer diagrams \cite{Franco:2005rj,Franco:2005sm} (see also \cite{Kennaway:2007tq,Yamazaki:2008bt} and references therein). However, the discussion of the deformed geometries has been carried out only for a few examples. In this paper we undertake the task of providing systematic recipes to build complex deformations of orientifolds of toric CY singularities, in terms of dimer diagram combinatorics. 

The basic strategy is to first characterize which toric singularities admit a given orientifold quotient, and then display the properties of complex deformations of the parent geometry required for them to survive in the orientifolded theory. This is most straightforwardly carried out in terms of the behaviour of zig-zag paths \cite{Hanany:2005ss,Feng:2005gw}, which correspond to external legs of the web diagram dual to the toric diagram of the singularity \cite{Aharony:1997ju,Aharony:1997bh,Leung:1997tw}. Complex deformations of the parent theory corrspond to removal of subsets of external legs in equilibrium (i.e. with zero total $(p,q)$ charge) \cite{Franco:2005zu}.

We anticipate our main results here. In order to do so, let us mention that as observed in \cite{Franco:2007ii}, orientifolds of toric singularities fall in two broad classes, those leaving fixed lines on the dimer and those with only fixed points. They have different geometric actions and thus need to be studied separately: 

$\bullet$ As we show, the criterion for a singularity to be compatible with fixed line orientifolds is that the web diagram accepts a $\mathbb{Z}_2$ action corresponding to a line reflection. Thus, if one wants to deform such a geometry, the resulting singularity after the complex deformation must also have a web diagram compatible with the same $\mathbb{Z}_2$ action. 

$\bullet$ The fixed point orientifold case  is quite different, and requires separating the set of all zig-zags into subsets such that all zig-zags in a subset have the same $(p,q)$ charge. The criterion for a singularity to be compatible with orientifold points is then that the number of such subsets with an odd   number of elements (zig-zags) is less than or equal to four. Finally, deformations of  singularities compatible with orientifold points require that the subweb in equilibrium removed from the web diagram must have an even number of zig-zags of each type, or phrased in another way, to remove two equal copies of a subweb in equilibrium. 

As final remarks, let us mention that in this paper we restrict ourselves to classical phases of dimers. Non-classical phases of dimers \cite{GarciaEtxebarria:2012qx,Garcia-Etxebarria:2013tba} have recently shown to be compatible with more orientifold actions \cite{Garcia-Etxebarria:2015hua}. 

 Also, it is important to note that our analysis includes recently found setups of toric singularities that do not accept resolutions if orientifolds are present \cite{Garcia-Etxebarria:2015lif}. The concrete setup of \cite{Garcia-Etxebarria:2015lif} involves a conifold with orientifold lines on its dimer.  Throughout the paper we will restrict to the analysis of toric CY's on singular and deformed phases, so we will not study the impossibility of having resolution phases. Let us just comment that this analysis is however straightforward for fixed line orientifolds: an orientifold of a singular geometry is incompatible with its resolutions if the corresponding singular web diagram admits no desingularization compatible with the $\mathbb{Z}_2$ symmetry.

The paper is organized as follows. In sections \ref{sec:dimer-general} we provide general background on  dimer technology and warped throats from toric CY singularities. In section \ref{sec:zigzag} we introduce a very useful tool for our analysis: zig-zag paths. Then, in section \ref{sec:dimer-mirror} we provide the mirror description of dimers and in \ref{sec:dimer-deform} we explain complex deformations on throats from this perspective. Finally, in section \ref{sec:dimer-orientifolds} we provide the necessary tools understand how orientifolds can act on dimers. In section \ref{sec:criterion-orientifolds} we start our analysis by giving criteria to tell which toric CY's are compatible with orientifolds on the dimer, first considering orientifold actions leaving  fixed lines on the dimer in section \ref{sec:criterion-lines}, and then those leaving fixed points in section \ref{sec:criterion-points}. Using this knowledge, in section \ref{sec:deform-orientifolds} we address the problem of which singularities are compatible with both orientifold actions and complex deformations. Once again, we address this issue first for singularities compatible with orientifold lines in section \ref{sec:deform-lines} and then for those with  orientifold points in section \ref{sec:deform-points}.

\bigskip

\section{Background} \label{sec:background}

In this section we provide general background on dimer models, starting with their basic features, their mirror description and afterwards review how complex deformations of toric singularities are described in this context. Finally, we mention the general features of orientifolds of dimer models that will be relevant in our analysis.

\bigskip

\subsection{Some generalities about dimers}
\label{sec:dimer-general}

\begin{figure}[htb]
\begin{center}
\includegraphics[scale=.25]{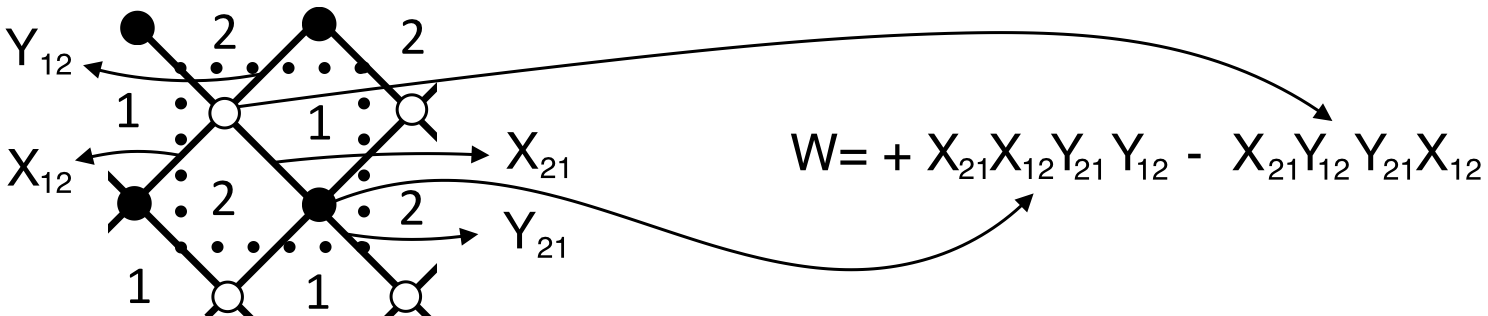} 
\caption{\small Dimer describing the conifold theory. Faces with labels 1 and 2 represent the two gauge factors $SU(n_1) $ and $ SU(n_2)$,  edges $X_{12}$ and $Y_{12}$ are chiral fields transforming in the   $( \bar{\mathbf{n}}_1,\mathbf{n}_2 )  $  representation of the gauge group whereas  edges $X_{21}$ and $Y_{21}$  transform in the   $( \mathbf{n}_1,\bar{\mathbf{n}}_2 )  $.   Finally, vertices/nodes are superpotential terms involving the fields touching the node. We take the convention that black nodes have plus sign in the superpotential and involve a product of the fields ordered in a clockwise direction, and white nodes have negative sign and involve a product of the fields ordered in a counter-clockwise direction. }
\label{fig:conifold-dimer}
\end{center}
\end{figure}

The  gauge theories arising from  D3-branes on toric CY singularities can be described in terms of bipartite tilings of $\mathbb{T}^2$ known as dimers or dimer diagrams \cite{Hanany:2005ve}. These diagrams consist of a series of faces, edges and vertices that translate to gauge groups, chiral multiplets  in bi-fundamental representations of the adjacent groups and superpotential terms respectively. Moreover, the fact that these singularities are toric results in vertices being of two types, the usual criterion being that we color them in black and white corresponding to e.g. plus and minus sign in the superpotential, and the edges of the dimer  always join vertices of different type. The  superpotentials of these theories thus include each bi-fundamental field in the theory twice in the superpotential, once with plus sign and once with minus sign \cite{Hanany:2005ve}. We provide an example showing the dimer of the conifold in figure \ref{fig:conifold-dimer}.

The usefulness of dimers relies on their powerful encoding of field theory phenomena into diagram combinatorics. One of the most interesting  is Seiberg Duality \cite{Seiberg:1994pq}, which allows to describe two UV field theories (related by a strong-weak duality) flowing to the same IR dynamics. This duality can be easily described in terms of  dimer diagrams \cite{Franco:2005rj} and is the reason why a unique singularity accepts different dimers, or equivalently  different toric phases: all the phases are related by a series of Seiberg dualities \cite{Beasley:2001zp,Feng:2001bn}. Moreover, these phases are important because the  RG flow of dimer theories in the presence of fractional branes (in the holographic picture these correspond to an anomaly-free increase of certain gauge groups) is described by a periodic series of Seiberg dualities, known as  duality cascade, that make the dimer of a given singularity go through its different toric phases. The RG flow was also shown to decrease the rank of all gauge groups  as the theory flows to the IR \cite{Klebanov:2000nc}. In terms of the supergravity dual, the RG flow corresponds to turning on  non-trivial  fluxes on the internal space such that the D3-brane charge decreases as one moves towards the singularity in the radial direction of the conical singularity, which corresponds to the decrease of the ranks of the gauge groups in the holographic picture.

Another important phenomenon that can be easily described in terms of dimers are complex deformations generalizing the Klebanov-Strassler  \cite{Klebanov:2000hb} smoothing of the conifold. In the supergravity description, complex deformations are the origin of the  fluxes we just mentioned: by putting $M$ units of RR 3-form flux on a 3-cycle to hold it at finite size, and a non-trivial flux of the NSNS 3-form flux along the dual 3-cycle, one finds that the D3-brane charge does indeed depend in the radial direction. The smoothing of the singularity by growing a 3-cycle changes the base of the conical singularity, which results in two different dimers describing the physics before and after the complex deformation. The relation between both dimers goes as follows: starting for the UV dimer, at certain energy level one or more gauge groups confine and the strong dynamics produces a quantum modification of the moduli space. The resulting theory corresponds to the IR dimer, describing the conical singularity after the deformation. One interesting feature of fluxed complex deformations is that on the gravitational side they give rise to warped throats \cite{Giddings:2001yu}, which as already mentioned are particularly interesting in order to create hierarchies. We will explain more about complex deformations on section \ref{sec:dimer-deform} after learning about the mirror of dimer theories.

The last useful property of dimers that we want to mention here is that they allow a diagrammatic representation of the effect of the orientifold on the gauge groups and matter content of the theory as described in \cite{Franco:2007ii}. Once again, we leave further details for section \ref{sec:dimer-orientifolds}.

\bigskip

\subsubsection{Zig-zag paths}
\label{sec:zigzag}

\begin{figure}[htb]
\begin{center}
\includegraphics[scale=.28]{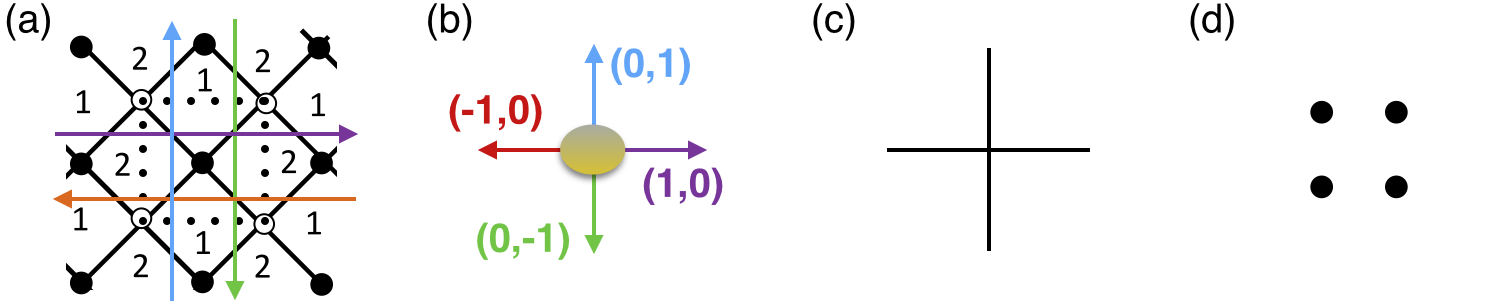} 
\caption{\small  (a) The unit cell of the conifold dimer together with its zig-zag paths. (b) The $(p,q)$'s of the zig-zag paths of the conifold as the external legs of its web diagram. (c) Web diagram of the conifold. (d) Toric diagram of the conifold, which is graph-dual to the web diagram.  }
\label{fig:conifold-zz}
\end{center}
\end{figure}

In this section we describe a tool that will be specially useful for our analysis, introduced in \cite{Hanany:2005ss} and known as zig-zags paths or simply zig-zags. These are oriented paths on the dimer crossing edges on the middle and turning maximally e.g. to the right at white nodes and to the left at black nodes. The different paths cross each other, but physically consistent graphs do not allow for paths intersecting themselves.   In figure \ref{fig:conifold-zz}(a) we show the dimer of the conifold together with its zig-zag paths. A crucial fact of zig-zags is that each path defines a homologically non-trivial 1-cycle on the dimer torus; once we define our unit cell for the dimer, each zig-zag has some associated winding numbers $(p,q)$. Note that zig-zags with the same  $(p,q)$ will be parallel on the dimer, and thus will also not cross each other. Different choices of unit cells related via $SL(2,\mathbb{Z})$ transformations imply corresponding changes on the $(p,q)$ vectors. 

It is of special relevance to us   the set of  $(p,q)$'s of all zig-zag paths of a given dimer, also known as the $(p,q)$ \textit{web} \cite{Hanany:2005ss}. This web defines the toric singularity  in a way that we now describe. The toric singularity is defined by a gauged linear sigma model (GLSM). The  D-terms of the GLSM can be encoded in a 2 dimensional diagram known as the \textit{web diagram} (for a review on toric geometry see e.g. \cite{Bouchard:2007ik}). The web diagram is a set of lines, which for our purposes can be split into two different types: internal and external lines (see figure \ref{fig:conifold-zz}(c) for an example). Internal lines describe Fayet-Iliopoulos terms on the GLSM, leading to resolutions of the singularity, but this is not relevant for our purposes.  External lines  are lines pointing at different directions, which can be characterized by a 2 dimensional vector. There is a one to one correspondence between the external lines of the web diagram and  the $(p,q)$'s of  zig-zag paths \cite{Hanany:2005ss}, such that a zig-zag path with winding numbers $(p,q)$      corresponds to an external leg of the web diagram pointing at the direction $(p,q)$ \cite{Hanany:2005ss}, see figures \ref{fig:conifold-zz}(b) and (c) for an example. 
 The CY condition on the GLSM then translates to  the sum of all $(p,q)$'s of a given singularity being zero, which is known as the web being in equilibrium. 
 It is important to note that the $(p,q)$ web, and thus web diagram, has many representatives related by $SL(2,\mathbb{Z})$ transformations, since the dimer also has different possible unit cells related by these transformations.  For future convenience, we define one more graph which also defines the toric singularity; this is the \textit{toric diagram}. The toric diagram is a 2 dimensional diagram which is   graph-dual to the \textit{web diagram}, see figures \ref{fig:conifold-zz}(c) and (d) for an example.  This diagram is a convex integer sublattice $Q\subset \mathbb{Z}^2$. For more relations between these diagrams and dimer diagrams we refer the reader to  \cite{Hanany:2005ss,Feng:2005gw}. 

\begin{figure}[htb]
\begin{center}
\includegraphics[scale=.28]{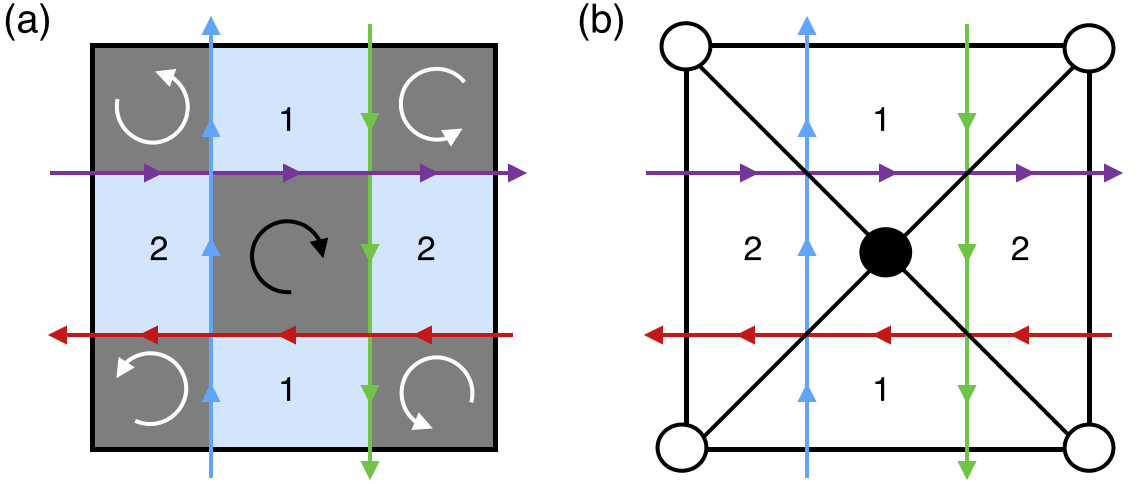} 
\caption{\small  Use of the fast forward algorithm to derive the dimer diagram of the conifold from the $(p,q)$ web. The $(p,q)$ web for the conifold is shown in figure\ref{fig:conifold-zz}(b).  (a) The zig-zag paths placed on the unit cell. They bound faces of two types: the ones shaded in grey are bound by zig-zag paths whose orientation always goes on the clockwise direction (black arrow) or counter-clockwise direction (white arrow); and the ones in light blue, bound by zig-zags that have \textit{opposite directions} at the points where they cross each other. (b) From the previous setup, we obtain the dimer by replacing faces in black by black and white nodes depending on the orientation of the zig-zags on the face,  zig-zag crossings map to edges bound by a white and a black node and the faces in blue correspond to gauge groups. }
\label{fig:algorithm}
\end{center}
\end{figure}

If one is interested instead in building up the dimer diagram from the toric data, zig-zag paths provide a nice recipe, which was dubbed the \textit{fast inverse algorith} in \cite{Hanany:2005ss}. In this section we provide an example describing how to obtain the conifold dimer from its $(p,q)$ web, and refer to \cite{Hanany:2005ss} for the general prescription. The $(p,q)$ web of the conifold was shown in figure \ref{fig:conifold-zz}(b). Starting with a square unit cell for the dimer for simplicity, the procedure lies in an ordered arrangement of zig-zags paths with windings $(p,q)$. For the conifold case, we illustrate in figure \ref{fig:algorithm}(a) how to arrange the zig-zags and in (b) how to obtain the corresponding dimer. 

\bigskip

\subsection{The mirror perspective}
\label{sec:dimer-mirror}

The mirror dual of a toric CY singularity  was shown in \cite{Feng:2005gw} to live in a threefold given by a double fibration over the complex plane
\beqa
 z & = & uv \nonumber \\
  z & = & P(x,y) = \sum_{m,n\in Q} c_{mn}x^my^n \label{eq:fibration}
\eeqa
where $u,v,z\in \mathbb{C}$ and $x,y\in \mathbb{C}^*$ are the coordinates defining the threefold and $P(x,y)$   is the Newton polynomial of the toric diagram of the singularity. In this picture\footnote{It actually corresponds to an intersecting brane configuration, in the sense of \cite{Aldazabal:2000cn,Aldazabal:2000dg}.}, the gauge groups of the dimer translate to D6-branes wrapping 3-cycles on the geometry, bifundamental fields  arise from open strings on the intersections between these branes and superpotential terms come from worldsheet instantons in discs bound by three or more branes. All intersections between branes, and thus all worldsheet instantons, were shown to meet at the Riemann surface  $\Sigma $ given by $P(x,y)=0$, on which we will focus from now on. For other aspects of the mirror dual we refer the reader to  \cite{Feng:2005gw}. 

\begin{figure}[htb]
\begin{center}
\includegraphics[scale=.3]{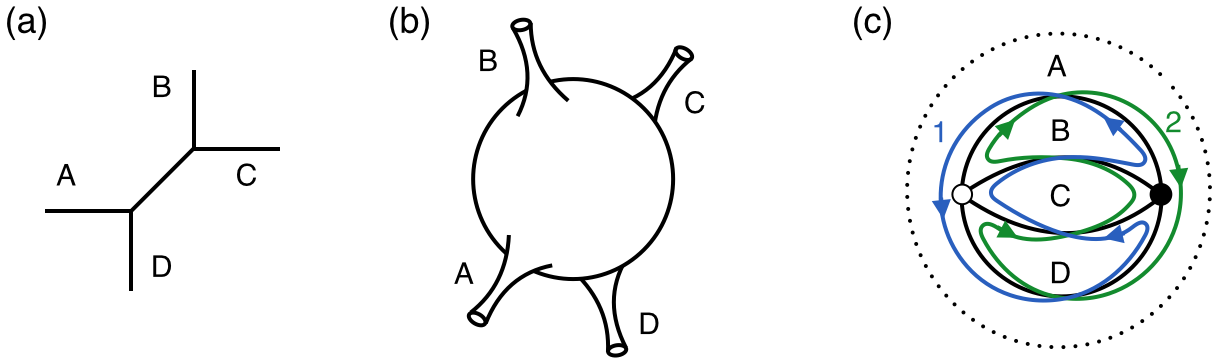} 
\caption{\small (a) Web diagram of the conifold on the resolved phase. Labels correspond to external legs of the diagram. (b) The curve $P(x,y)=0$ of the conifold is a thickening of its web diagram. (c) The tiling of  $\Sigma $ for the conifold. The dotted line corresponds to a unique point, since $\Sigma$ is a sphere with punctures as shown in (b), corresponding to faces in (c). Also, comparing with figure \ref{fig:conifold-zz}(a) we see that it has four edges and one vertex of each color. The paths in green and  blue are the D6-branes giving rise to the two gauge groups.  }
\label{fig:shiver}
\end{center}
\end{figure}

The surface $\Sigma $, defined by $P(x,y)=0$, is a Riemann surface with handles and punctures.  This surface was shown in \cite{Feng:2005gw} to be a thickening of the web diagram \cite{Aharony:1997bh,Aharony:1997ju,Leung:1997tw}, with punctures corresponding to the external legs of the web diagram, and its genus $g$ is the same as the number of internal points of the toric diagram, which as said before is graph dual to the web diagram. In figures \ref{fig:shiver}(a) and (b) we show the web diagram and the curve $\Sigma$ corresponding to the conifold. The D6-branes giving rise to gauge groups wrap 1-cycles in $\Sigma$ surrounding some of its punctures and intersecting each other.  Open strings  on these intersections give the bifundamental chiral fields and worldsheet instantons on the discs bound by several D6-branes and their intersections are responsible for superpotential terms. The way D6-branes are wrapping 1-cycles in $\Sigma$ so that they give rise to the same chiral fields and superpotential terms as those in the dimer was described in \cite{Feng:2005gw}.  The  outcome is that D6-branes are wrapped such that each puncture in $\Sigma$ is surrounded by a series of D6-brane intersections and worldsheet instantons, or equivalently fields and superpotential terms. 

It is thus possible to define a bipartite graph tiling on $\Sigma$, where each face represents a puncture of $\Sigma$, or equivalently an external leg of a web diagram or zig-zag of the dimer. Edges of this tiling are the fields in the intersection(s) between zig-zag paths in the dimer, and their vertices correspond to the instantons (superpotential terms), which we can color as in the dimer, since each edge must have a white and a black node on each side. Finally, the 1-cycles wrapped by D6-branes are zig-zag paths of this tiling of $\Sigma$, i.e. turning maximally when they are next to a black node and maximally to the left when the node is white. We thus have two tilings, the original dimer and the one we just described, which are strongly related via the so called untwisting procedure in  \cite{Feng:2005gw} and with two main differences: on the one hand, the dimer is defined on a torus, whereas the tiling in the mirror has genus $g$. On the other hand, in the original dimer faces represented gauge groups while zig-zag paths corresponded to external legs of the web diagram, whereas in the mirror, faces are the external legs and zig-zags are D6-branes corresponding to gauge groups. In order to avoid possible confusions, from here on we will use the term dimer only to describe the tiling of the torus where faces represent gauge groups, and zig-zag paths will be the paths we defined on the dimer; we will not use these terms for the tiling of the mirror curve $\Sigma$.

\bigskip

\subsection{Complex deformations}
\label{sec:dimer-deform}

One of the main motivations to be interested on toric CY singularities is that they provide an interesting scenario to create hierarchies \cite{Giddings:2001yu}. On the gravity side these are created by the warping of the internal manifold, after turning on certain fluxes that hold a 3-cycle  at finite size on the internal manifold, leading to geometries known as warped throats. The first known  manifold of this kind is the deformed conifold in \cite{Klebanov:2000hb}, but nowadays there are many examples of this kind on the market, see e.g.  \cite{Franco:2005fd,Retolaza:2015sta,Franco:2015kfa,Retolaza:2015nvh}. The growth of the 3-cycle or complex deformation can be understood from different perspectives;  we are interested in the description of such phenomenon in terms of the web diagram, gauge theory and the mirror geometry. Readers interested in further details on the gravitational picture can find them in e.g. \cite{Klebanov:2000hb,Franco:2005fd}. 

\begin{figure}[htb]
\begin{center}
\includegraphics[scale=.27]{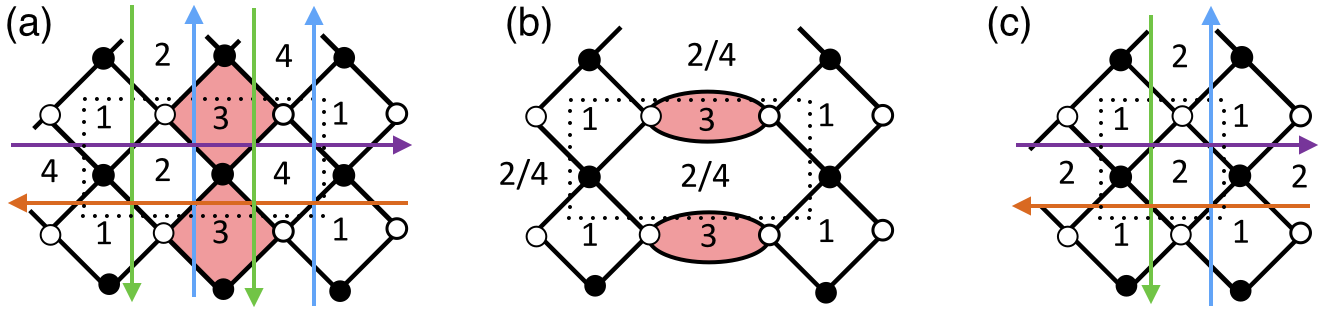} 
\caption{\small (a) Dimer diagram of the $\mathbb{Z}_2$ orbifold of the conifold describing the UV physics. Gauge group with label 3 is taken to have $N_f\leq N_c$ and thus confines. (b) An intermediate step in the confinement/deformation process following the recipe in \cite{Franco:2005fd}. (c) The resulting dimer after the deformation process, that describes the IR physics of the gauge theory. This dimer corresponds to the conifold. }
\label{fig:conifz2-a}
\end{center}
\end{figure}

As already mentioned, the gauge theory description of complex deformations can be easily carried out using dimer diagrams \cite{Franco:2005zu}. By placing fractional branes on some gauge groups (we will shortly give a criterion to determine which) we increase their ranks and break conformal invariance. This triggers a cascade of Seiberg dualities that periodically reduces the rank of all gauge groups by a unique number that depends on how many fractional branes we put \cite{Klebanov:2000hb,Franco:2005fd}. At certain point on the RG flow the theory reaches a point where the groups with fractional branes have number of colors and flavours satisfying $N_f\leq N_c$, so their strong dynamics  leads  to a modification of the moduli space and thus modification of the dimer. The resulting gauge theory and dimer have less gauge factors and correspond to the left-over (possibly singular) geometry after the complex deformation. Therefore, the singularity and thus the theory are different before the complex deformation (UV of the gauge theory) and after it (IR of the gauge theory). We show an example of this process in figure  \ref{fig:conifz2-a}.

\begin{figure}[htb]
\begin{center}
\includegraphics[scale=.2]{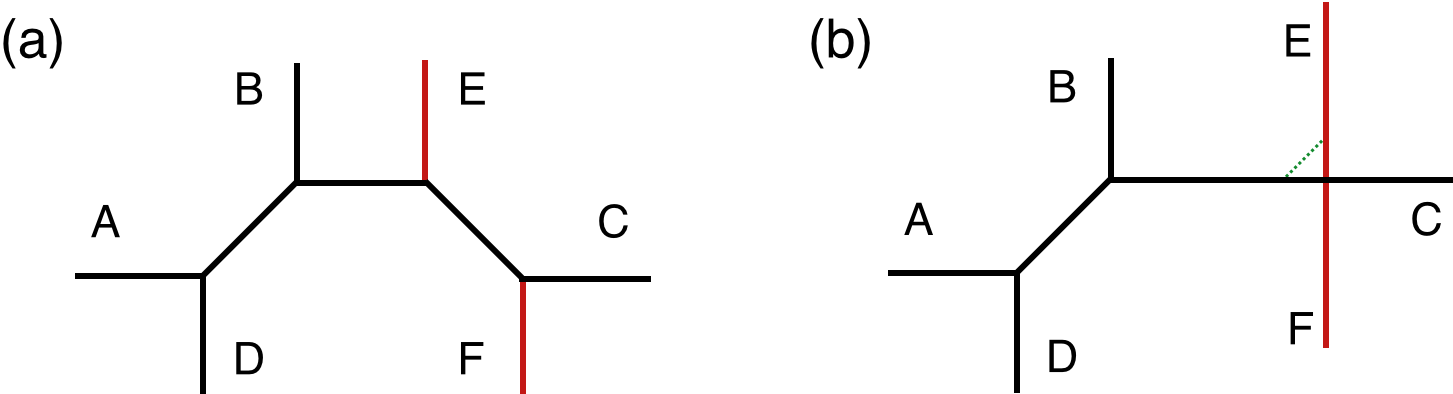} 
\caption{\small (a) Web diagram for the $\mathbb{Z}_2$ orbifold of the conifold on the resolved phase. (b) Web diagram representation of the complex deformation from the $\mathbb{Z}_2$ orbifold of the conifold to the conifold. The dashed green line represents the separation between the conifold web diagram and the subweb in equilibrium we removed. The dashed line also represents pictorially the 3-cycle grown in the deformation process.  }
\label{fig:conifz2-b}
\end{center}
\end{figure}

The  difference between the UV and the IR theories is also reflected  on the web  diagram, where the deformation process corresponds to the removal of a sub-web in equilibrium. After this removal we are left with a new web diagram also in equilibrium and a smaller toric diagram of smaller area, in agreement with having less gauge factors in the IR (the area of the toric diagram was shown in \cite{Hanany:2005ss} to be twice the amount of gauge factors on the dimer). The groups where we placed the confining fractional branes are those bound by the zig-zag paths corresponding to the external legs that have been removed. We illustrate this in figure \ref{fig:conifz2-b}(a). 

\begin{figure}[htb]
\begin{center}
\includegraphics[scale=.25]{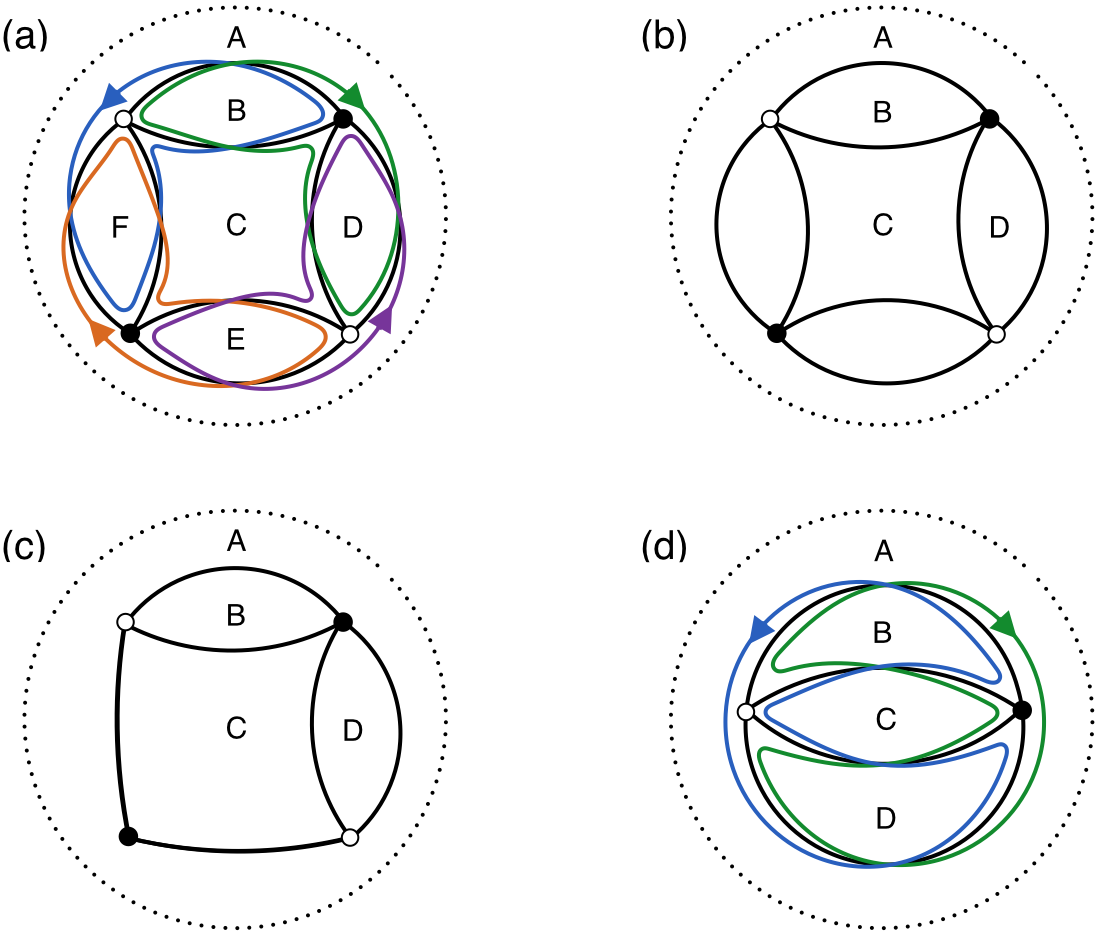} 
\caption{\small (a) Tiling of the mirror surface $\Sigma$ of the $\mathbb{Z}_2$ orbifold of the conifold. The paths in colours represent the four D6-branes giving rise to the gauge groups in the gauge theory. The confining group is represented by the D6-brane in orange. (b) Tiling of $\Sigma$ after cutting out the tiles corresponding to external legs E $\&$ F. (c) Tiling  of $\Sigma$ after gluing together the boundaries left after cutting out the tiles. The black node touching only two edges represents a mass term for the two fields. (d) The tiling of $\Sigma$ after integrating out the massive fields together with the D6-branes left after the deformation. We see that the D6-branes in blue and purple in (a) now recombined to the one we draw in blue. This $\Sigma$ corresponds to the conifold. }
\label{fig:conifz2-c}
\end{center}
\end{figure}

Finally, since the surface $\Sigma$ is a thickening of the web diagram, the deformation process  must necessarily change this surface as  explained in  \cite{GarciaEtxebarria:2006aq}. In the mirror surface $\Sigma$ the D6-branes corresponding to the confining gauge groups  wrap certain punctures, which correspond to the external legs of the web diagram that we will remove. The removal of the external legs corresponds to cutting out these faces from $\Sigma$ and then gluing together the boundaries of the  surface left after cutting them. This process involves a recombination of the remaining D6-branes on the mirror, corresponding to the higgsing in the gauge theory due to mesons that get a \textit{vev}.  The new surface we are left with is the one describing the IR physics. Figure \ref{fig:conifz2-c} shows an example of this kind.

\bigskip

\subsection{Orientifolds of dimer models}
\label{sec:dimer-orientifolds}

The last object to be discussed before moving on to new ideas are orientifolds of dimers. These were first studied in  \cite{Franco:2007ii} and here we  only summarize the relevant features for the following sections.

Orientifolds are the key ingredient to eliminate some  degrees of freedom of a theory such that the outcome is a theory with different gauge factors and matter representations. Regular dimers only have gauge groups of the $SU(N)$ type and matter in bifundamental and adjoint representations. When an orientifold action is implemented on a toric CY singularity, the theory can also have $SO(N)$ and $USp(N)$ types of gauge groups depending on the orientifold charges, and also matter in the two index symmetric and antisymmetric representations. The way to obtain these new degrees of freedom in dimers was described in \cite{Franco:2007ii}.

  Dimer diagrams live on tori, and this restricts the possible orientifold actions they accept. The dimer involutions that we focus on in this paper are those studied in \cite{Franco:2007ii}, which correspond to $\mathbb{Z}_2$ actions leaving fixed loci on the dimer (other $\mathbb{Z}_2$ actions that do not have fixed loci are possible but were not studied in the literature so far). These can be of  two different types: those leaving fixed lines, also known as orientifold lines, and those leaving fixed points, known as orientifold points. The geometric action of the two types of orientifolds is different, and thus they act in a different manner on the mesonic operators (gauge invariant operators from field products of the theory). In particular, they act in a different way on superpotential terms,  a property that  makes it better to  study them separately.  

\begin{figure}[htb]
\begin{center}
\includegraphics[scale=.27]{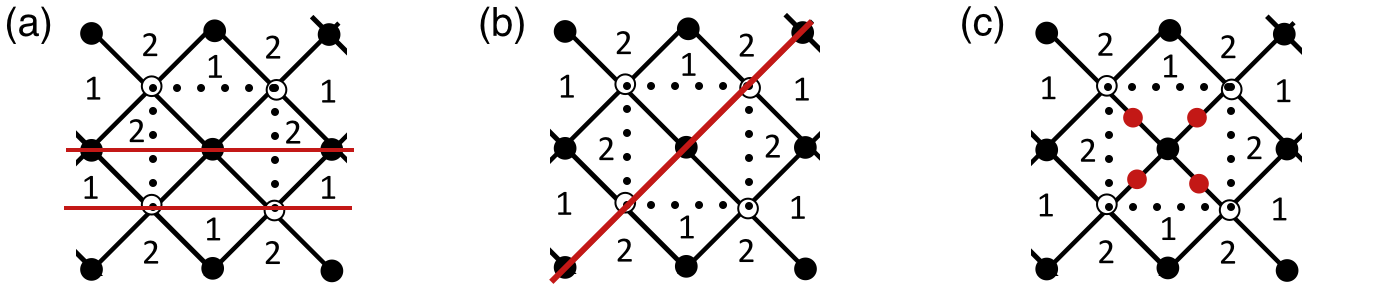} 
\caption{\small (a) Dimer of the conifold with  orientifold lines inverting one of the coordinates. It relates nodes with the same colour between them. (b) Dimer of the conifold with an orientifold line exchanging the coordinates. It also relates nodes with the same colour between them. (c) Dimer of the conifold with   orientifold points. This type of orientifold maps nodes of different colour. }
\label{fig:orientifolds}
\end{center}
\end{figure}

We will first deal with orientifold lines.  These can be of two types as shown in figures \ref{fig:orientifolds}(a) and (b) depending on whether there are two parallel orientifold lines, or a single one.
Since the orientifold line can be thought of as a boundary of the orientifold of the dimer, we can say that these dimers live on a cylinder and a M\"{o}bius strip, respectively. 
These kinds of orientifolds were shown to relate superpotential terms corresponding to vertices with same color, as can be seen in the examples of figures \ref{fig:orientifolds}(a) and (b). 

In the case of orientifold points the geometric action inverts both coordinates of the $\mathbb{T}^2$, with four fixed points, as shown in figure \ref{fig:orientifolds}(c).   This sets a series of identifications that implies that dimers with orientifold points live in a sphere, as we will prove in section \ref{sec:criterion-points-mirror}. 
This time the orientifold relates vertices/nodes with different color.
This sets some restrictions, since this is only possible if orientifold points fall on top of edges of the dimer or in the middle of hexagonal faces.

\bigskip

\section{Toric singularities compatible with orientifolds}
\label{sec:criterion-orientifolds}

In order to find out which kind of singularities are compatible with both complex deformations and orientifold actions, we will first provide criteria to easily guess if a singularity is compatible with orientifold actions, and if so, which kind(s) of orientifold action(s) it admits. We anticipate that all the information is encoded on the set of all \textit{zig-zag paths} of the dimer, or equivalently the external legs of the  \textit{web diagram}, and thus the \textit{toric diagram}.

\bigskip

\subsection{Singularities compatible with orientifold lines}
\label{sec:criterion-lines}

Our strategy to explain which toric CY's can have orientifold actions leaving fixed lines on the corresponding dimer will be to start considering an example of this kind and derive the general rules from it. 

As explained in section \ref{sec:dimer-orientifolds}, dimers with orientifold lines can be of two kinds. 
 Both cases have many common features, so it is enough to study one of these cases and then extrapolate the knowledge to the other case.

\begin{figure}[htb]
\begin{center}
\includegraphics[scale=.2]{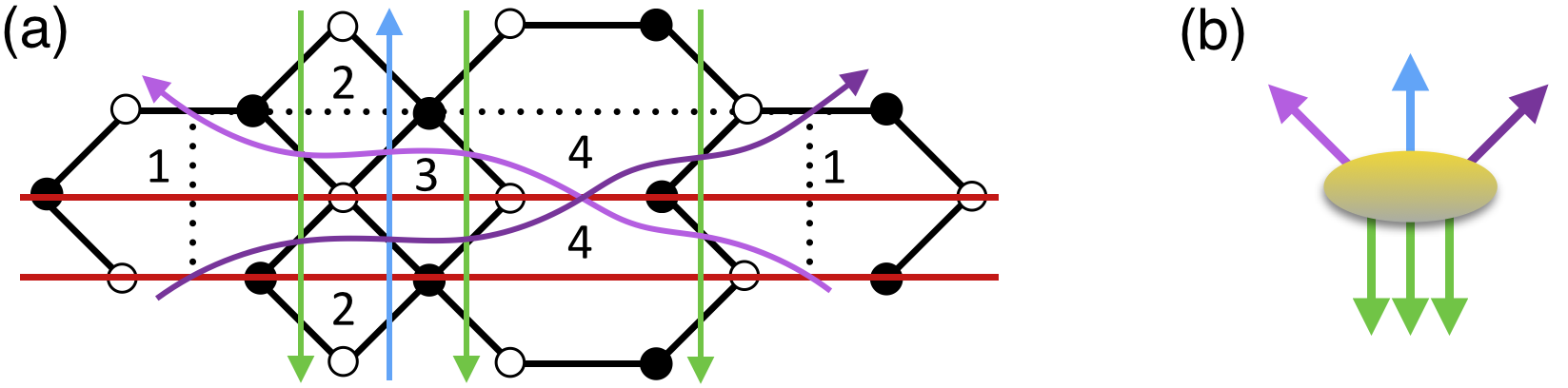} 
\caption{\small  (a) Unit cell of the dimer for $L^{1,3,1}$ with orientifold lines in red, and zig-zag paths in colors; those with the same winding numbers $(p,q)$ have the same colour. (b) Set of $(p,q)$'s of the zig-zag paths of $L^{1,3,1}$ as the external legs of its web diagram. }
\label{fig:L131-dimer}
\end{center}
\end{figure}

The example we will use is the dimer of the singularity known as $L^{1,3,1}$ \cite{Benvenuti:2005ja,Butti:2005sw}, that was shown in \cite{Franco:2007ii} to accept orientifold actions with parallel fixed lines. The corresponding dimer and  its zig-zag paths are shown in figure \ref{fig:L131-dimer}.

\begin{figure}[htb]
\begin{center}
\includegraphics[scale=.25]{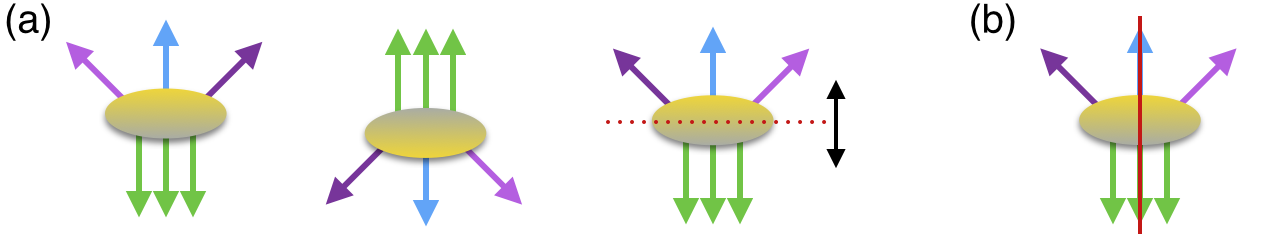} 
\caption{\small (a) Starting with the external legs of the web diagram of $L^{1,3,1}$ in \ref{fig:L131-dimer}(b), if we first invert all zig-zags and then act on them with the geometric action of the orientifold line on the dimer, i.e. invert the vertical direction, we end up with a web diagram that looks  the same as the initial one, up to certain identifications between different zig-zag paths that agree with the identifications on the dimer of figure \ref{fig:L131-dimer}(a). (b) The $\mathbb{Z}_2$ action that leaves a line invariant on the web diagram of $L^{1,3,1}$ and is equivalent to doing the two previous steps. This time it corresponds to a vertical line. }
\label{fig:L131-legs}
\end{center}
\end{figure}

From figure  \ref{fig:L131-dimer}(a) we observe that the  zig-zag paths of $L^{1,3,1}$ are compatible with the $\mathbb{Z}_2$ symmetry of the orientifold: after inverting the direction of any zig-zag path, it becomes the orientifold image of one of the original zig-zag paths.  
This symmetry can be made more explicit in a two step process. First, one must invert the $(p,q)$ winding numbers of all zig-zags to reverse their orientation
\beq
\left\lbrace \left( \begin{matrix}
p_i \\ q_i \end{matrix} \right)\right\rbrace_{i=1, ..., Z}\quad \rightarrow \quad \left\lbrace \left( \begin{matrix}
-p_i \\ -q_i \end{matrix} \right)\right\rbrace_{i=1, ..., Z}\ . \label{eq:inversion}
\eeq
$Z$ is the number of zig-zags of the singularity.  Since our orientifold inverts the vertical component of the dimer, we must do the same with the inverted $(p,q)$'s: 
\beq
\left\lbrace \left( \begin{matrix}
-p_i \\ -q_i \end{matrix} \right)\right\rbrace_{i=1, ..., Z}\quad \rightarrow \quad \left\lbrace \left( \begin{matrix}
-p_i \\ q_i \end{matrix} \right)\right\rbrace_{i=1, ..., Z} .
\eeq
Since there exists a $\mathbb{Z}_2$ symmetry allowing for an orientifold of our singularity, the resulting set of zig-zag paths must agree with the original one:
\beq
\Omega : \  \left\lbrace \left( \begin{matrix}
p_i \\ q_i \end{matrix} \right)\right\rbrace_{i=1, ..., Z}\quad \rightarrow \quad \left\lbrace \left( \begin{matrix}
-p_i \\ q_i \end{matrix} \right)\right\rbrace_{i=1, ..., Z}= \left\lbrace \left( \begin{matrix}
p_j \\ q_j \end{matrix} \right)\right\rbrace_{j=1, ..., Z}\ . \label{eq:pq-line}
\eeq
 In our case, we see that those paths without a horizontal component remain the same after the two steps and thus are their own orientifold images, whereas for zig-zags winding the horizontal 1-cycle  the orientifold relates two different zig-zags. We can follow this process diagrammatically by starting with the external legs of the web diagram and doing the same inversions. The process is shown in figure \ref{fig:L131-legs}(a). Comparing the initial and final web diagrams in figure \ref{fig:L131-legs}(a)  we  see that  zig-zags in green and blue are their own orientifold images, and the ones in light and dark purple are the image of one another, in agreement with figure \ref{fig:L131-dimer}(a). 

In fact, since in our case we first inverted  both components of the $(p,q)$'s and then we inverted again the vertical one, we see that the outcome is just inverting the horizontal component, and thus the set of external legs in figure \ref{fig:L131-legs}(a)  are related by a $\mathbb{Z}_2$ action that leaves invariant a vertical line, which is shown in figure \ref{fig:L131-legs}(b). Moreover, since physics is independent of our choice of what is horizontal and what is vertical on the dimer, the  idea to keep in mind is that  the external legs of the web diagram are compatible with a $\mathbb{Z}_2$ action leaving a horizontal/vertical line invariant because the dimer is compatible with  parallel orientifold lines. Indeed, this idea is general and works for any toric singularity. Moreover we can use this property the other way around to state that for any singularity whose external legs accept a $\mathbb{Z}_2$ action that leaves a horizontal/vertical line invariant, the corresponding dimer accepts orientifold lines.

\bigskip

\begin{figure}[htb]
\begin{center}
\includegraphics[scale=.3]{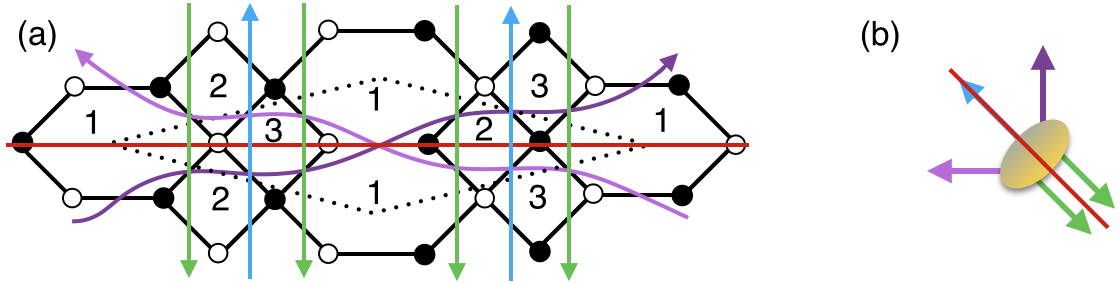} 
\caption{\small (a) The unit cell of the dimer of SPP, together with the orientifold line crossing the diagonal of the cell and the zig-zag paths.  (b) External legs of the web diagram of  SPP. The line in red is the fixed line left invariant under the   $\mathbb{Z}_2$ action of the orientifold line. Since the orientifold line crosses the dimer in the diagonal direction of the unit cell, the red line has a diagonal direction. }
\label{fig:SPP-line}
\end{center}
\end{figure}

The next case to consider is that of dimers with a single fixed line. In figure \ref{fig:SPP-line}(a) we show the dimer corresponding to the Suspended Pinched Point (SPP) as well as its zig-zag paths. This time the fixed line under the orientifold action exchanges both coordinates.
The corresponding $(p,q)$'s are shown in figure 	\ref{fig:SPP-line}(b) together with the orientifold line action leaving the web diagram invariant.

Following the same procedure as before, the identification of zig-zags with their orientifold images can be done in a two step process. We first invert all the winding numbers of zig-zag paths as  in (\ref{eq:inversion}), and then, since the fixed line is diagonal and its action on the dimer is to identify the two 1-cycles, we must act on the $(p,q)$'s by exchanging the two entries. Since this $\mathbb{Z}_2$ action is a symmetry of the dimer, the set of $(p,q)$'s we are left with is the same we had in the beginning. 
\beq
\Omega : \  \left\lbrace \left( \begin{matrix}
p_i \\ q_i \end{matrix} \right)\right\rbrace_{i=1, ..., Z}\quad \rightarrow \quad \left\lbrace \left( \begin{matrix}
-q_i \\ -p_i \end{matrix} \right)\right\rbrace_{i=1, ..., Z}= \left\lbrace \left( \begin{matrix}
p_j \\ q_j \end{matrix} \right)\right\rbrace_{j=1, ..., Z}\ . \label{eq:pq-line-diag}
\eeq
Once again, the $\mathbb{Z}_2$ symmetry is clearly manifest on the web diagram, as can be seen on figure \ref{fig:SPP-line}(b). The difference with the previous case is that for $L^{1,3,1}$ the line invariant under the $\mathbb{Z}_2$ action on the web diagram was horizontal, whereas now we have a diagonal one. It is a general fact that singularities compatible with diagonal orientifold lines have web diagrams that are $\mathbb{Z}_2$ symmetric with respect to a fixed line. 

\bigskip

Knowing how to characterize the action of orientifold lines on the web diagrams, we can use the criterion in the inverse direction to know which singularities are compatible with this type of action:

\begin{flushleft}
\noindent
\textbf{\underline{Criterion for toric CY singularities accepting orientifold lines:}} A toric CY singularity can have orientifold lines on its dimer if its web diagram has a $\mathbb{Z}_2$ symmetry that leaves a line invariant. Moreover, if the fixed line on the web diagram is horizontal/vertical, the orientifold line will invert one coordinate on the dimer; whereas for diagonal fixed lines on the web diagram, it will exchange its two coordinates. 
\end{flushleft}

The fact that a singularity admits a dimer with orientifold lines does not mean that all its toric phases are compatible with it. Starting from a phase compatible with this action, only the phases related to the initial one by Seiberg dualities performed in a $\mathbb{Z}_2$ symmetric way will also be compatible with the orientifold line. In general, there are also other Seiberg dualities which are not symmetric with respect to the fixed line action and thus lead to phases where the orientifold line action is not possible. 

In order to build up toric phases which are compatible with orientifold lines, one can   use  the \textit{fast inverse algorithm} in \cite{Hanany:2005ss} while  taking into account the required $\mathbb{Z}_2$ symmetry. We recommend to start  from a square unit cell together with the fixed line(s) that cross(es) it, in a direction fixed by the $(p,q)$ web in the sense of the criterion above. The easiest way to start placing zig-zag paths is usually by first putting the pairs that are paired between themselves and cross each other, since they cross each other on top of the fixed line(s) and thus they may be more difficult to place in later steps. The next convenient step is to include the zig-zags mapped to themselves, as these also cross the fixed line. Finally, the remaining paths are the ones mapped between themselves but not crossing each other, since they do not touch the orientifold line.

Finally, we comment on the different representatives of the $(p,q)$ web and thus the unit cell, since the  $\mathbb{Z}_2$ symmetry may or may not be manifest on certain representatives, as we now explain.

\subsubsection*{A remark on modular invariance of the unit cell}
\label{sec:modular}

\begin{figure}[htb]
\begin{center}
\hspace*{5pt}\includegraphics[scale=.27]{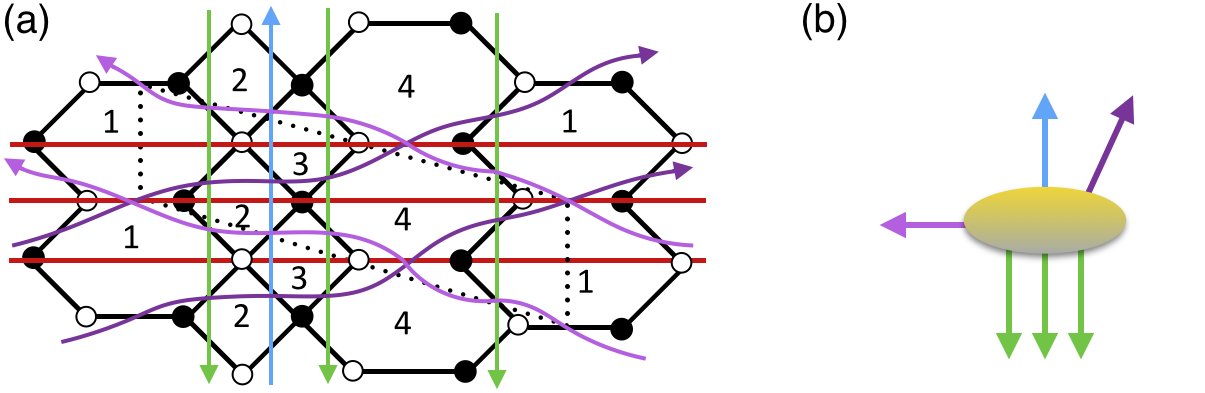} 
\caption{\small (a) Dimer for $L^{1,3,1}$ with orientifold lines and a different unit cell to the one in figure \ref{fig:L131-dimer}. (b) External legs of the web diagram of $L^{1,3,1}$ for the unit cell in (a). This unit cell does not make manifest on the web diagram the the $\mathbb{Z}_2$ symmetry required to have orientifold lines. }
\label{fig:L131-dimer-2}
\end{center}
\end{figure}

The fact that dimers (without orientifolds) are defined on tori allows for different unit cells related via $SL(2,\mathbb{Z})$ transformations. Different choices of unit cell thus lead to changes on the $(p,q)$ winding numbers of the zig-zag paths, and thus to different representatives of the web and toric diagrams. In the previous examples, the unit cell was chosen to make the $\mathbb{Z}_2$ symmetry obvious on the web diagram, this can be seen e.g. in figure \ref{fig:L131-legs}. It is important to understand the general situation, so we consider the $L^{1,3,1}$ theory with a different choice of unit cell, which is shown in  figure \ref{fig:L131-dimer-2}(a). The $(p,q)$'s of the zig-zag paths are now different as can be seen in figure \ref{fig:L131-dimer-2}(b), where it is not manifest that the singularity is compatible with orientifold lines. 

Since the change of unit cell is just a $SL(2,\mathbb{Z})$ transformation acting on the set of  all $(p,q)$'s, we can find out if a singularity allows orientifold lines just by checking if the whole set is invariant under the action of a $GL(2,\mathbb{Z})$ matrix in a way we now explain. 

We start by noting that for e.g. $L^{1,3,1}$, in (\ref{eq:pq-line}) the action of the orientifold on the $(p,q)$'s can be represented by a matrix, that in our  case was given by  $ \diag (-1,1) $. 
\beq
\Omega : \ \left\lbrace \left( \begin{matrix}
p_i \\ q_i \end{matrix} \right)\right\rbrace_{i=1, ..., Z} \! \! \! \! \!  \rightarrow  \ \left\lbrace  \left( \begin{matrix}
-1 & 0 \\ 0 & 1 \end{matrix} \right)    \left( \begin{matrix}
p_i \\ q_i \end{matrix} \right)\right\rbrace_{i=1, ..., Z}\! \! \! \! \! =\left\lbrace \left( \begin{matrix}
-p_i \\ q_i \end{matrix} \right)\right\rbrace_{i=1, ..., Z}\! \! \! \! \!  =\left\lbrace \left( \begin{matrix}
p_j \\ q_j \end{matrix} \right)\right\rbrace_{j=1, ..., Z} \! \! \!   \label{eq:pq-line-general}
\eeq
A change in the unit cell  changes this matrix to a more general $M_\Omega$:
\beq
\left( \begin{matrix}
-1 & 0 \\ 0 & 1 \end{matrix} \right) \quad \rightarrow \quad M_\Omega = \pm M^{-1} \left( \begin{matrix}
-1 & 0 \\ 0 & 1 \end{matrix} \right) M \quad , \quad M\in SL(2,\mathbb{Z}) \ , \label{eq:matrix1}
\eeq
where the two possible signs stand for the different choices of what we consider the horizontal  and the vertical axis. Therefore, we conclude that  for a general representative of the toric/web diagram, if the set of all $(p,q)$'s is invariant under the action of a matrix $M_\Omega$ given by (\ref{eq:matrix1}), then this singularity accepts two parallel orientifold lines. 

\bigskip

For the case of dimers with a single diagonal orientifold line, we can do the same and find the type of matrices that  leave the set of $(p,q)$'s for a singularity accepting such action on the dimer. The simplest matrix can be easily read from 
 (\ref{eq:inversion}). This time, we see that the matrices are of the kind
 \beq
M_\Omega =\pm  M^{-1}\left(\begin{matrix}
0&  1 \\  1 & 0 \end{matrix} \right) M \quad , \quad M\in SL(2,\mathbb{Z})\ , \label{eq:matrix2}
\eeq
where the sign once again depends on our choice of what is the horizontal axes, and thus is irrelevant for the physics.

\bigskip

Even though this analysis is more general than the one based on the $\mathbb{Z}_2$ symmetry of certain representative of the web diagram, if we want to find out if a toric CY singularity accepts orientifold lines, finding a matrix that leaves the set of $(p,q)$'s invariant is not a very practical approach. Instead, we recommend to look for the most symmetric representative of the toric/web diagram, where this symmetry is clearly manifest.

\bigskip

\subsubsection*{The mirror perspective}
\label{sec:criterion-lines-mirror}

We now turn to explain the effect of the orientifold on the dual mirror theory. This perspective is not necessary for the following analysis, so the uninterested reader may skip it. Nonetheless, we find it quite visual and it can be helpful specially  in section \ref{sec:deform-orientifolds}, since complex deformations of singularities are well understood on the mirror \cite{GarciaEtxebarria:2006aq}. 

As explained in \cite{Franco:2007ii}, in the mirror picture the Type IIB orientifolds map to an O6-plane in Type IIA. This O6-plane  preserves the same supersymmetry as the D6-branes giving rise to the gauge groups, and it usually intersects the punctured Riemann surface $\Sigma$ given by $P(x,y)=0$ in (\ref{eq:fibration})\footnote{There are cases where the O6 plane does not intersect this surface that we will study on section \ref{sec:criterion-points-mirror}. }. Starting from the original surface $\Sigma$, the theory with the orientifold will have a new mirror dual that  lives on a new Riemann surface $\Sigma '$. This new surface $\Sigma '$ is related to the initial one  as follows.  In the simplest situations, the O6-plane will wind a 1-cycle in $\Sigma$ that splits this surface into two surfaces $\Sigma '$ with  boundaries. The boundary is the  1-cycle wrapped by the O6-plane on the original surface, and the two surfaces are orientifold images of one another.   Slightly more involved situations are those involving  cross-caps on $\Sigma '$, where the mirror of the orientifolded theory does not split $\Sigma$ into two surfaces, but is given by a unique surface $\Sigma '$ that may or may not have boundaries (intersections with the O6).

\begin{figure}[htb]
\begin{center}
\hspace*{5pt}\includegraphics[scale=.23]{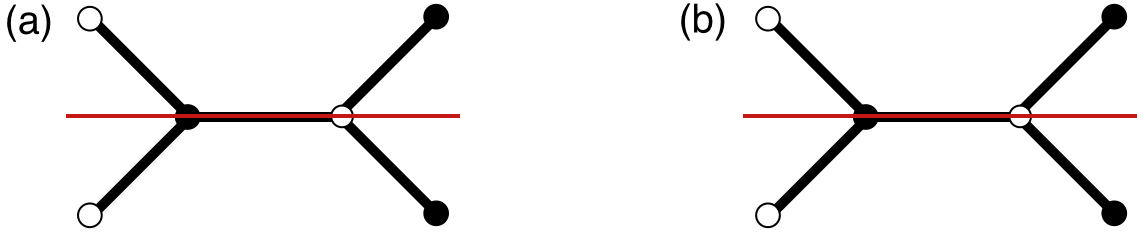} 
\caption{\small (a) Part of a dimer with an orientifold line where we see that certain fields and superpotential terms fall on top of the orientifold. (b) In the mirror of the previous setup we find that the same fields and superpotential terms fall on top of the boundary of $\Sigma '$, this time the difference being the meaning of the adjacent tiles that are zig-zag paths instead of gauge groups.}
\label{fig:boundary-lines}
\end{center}
\end{figure}

Our starting point will be to consider a  generic dimer where the numbers of gauge groups, fields, superpotential terms and zig-zag paths before the orientifold are $G,\ F,\ T$ and $Z$ respectively. The corresponding  numbers after the orientifold identification are   $G',\ F',\ T'$ and $Z'$, and the numbers of elements mapped to themselves  $n_G$, $n_F$, $n_T$ and $n_Z$. The relation between the numbers before and after the orientifold are
\beq
G'=\dfrac{G+n_G}{2} \quad ; \quad  F'=\dfrac{ F+ n_F}{2} \ \ \nn
\eeq
\beq
T'=\dfrac{T+n_T}{2} \quad ; \quad  Z'=\dfrac{ Z+ n_Z}{2}  \ . \label{eq:all-relations}
\eeq
By looking at e.g. the dimer of figure \ref{fig:L131-dimer}(a) we observe that the number $n_T$ of  superpotential terms mapped to themselves equals the sum $n_G+n_F$ of  gauge groups and  fields mapped to themselves. This is a general feature of orientifold lines. A similar relationship can be easily obtained on the mirror surface by observing that the fields and superpotential terms mapped to themselves must do the same on the tiling of  $\Sigma '$, see figure \ref{fig:boundary-lines}. 

The  difference on the mirror is that the faces represent zig-zags, so in this case there is one superpotential term per each field and each zig-zag. Putting both things together we see that
\beq
n_T=n_G+n_F=n_Z+n_F \quad \rightarrow \quad n_Z=n_G \ , \label{eq:nrelations}
\eeq
so the  numbers of gauge groups $n_G$ and zig-zags $n_Z$ that are mapped to themselves is the same when we have orientifold lines. These fields and terms falling on top of the orientifold  are relevant for other things. For example, it is important not to count them when computing the Euler characteristic of the surface where the orientifold theories live. Therefore, we  define the Euler characteristic of the orientifold of the dimer and the orientifold of the mirror this way:
\beq
\chi (\mathbb{T}^2 /\Omega ) = G'+T'-F' -(n_T-n_F) \ ,
\eeq
\beq
\chi (\Sigma ') = Z'+T'-F' -(n_T-n_F) \ .
\eeq

Using these expressions and (\ref{eq:nrelations}) it is easy to check that the equation
\beq
\chi (\Sigma ' \equiv \Sigma /\Omega)= \dfrac{\chi (\Sigma )}{2} \label{eq:euler}
\eeq
between the Euler characteristic of a Riemann surface and its orientifold surface holds for singularities whose dimers can have orientifold lines.
We can use this equation to relate  the genus $g$ of the initial surface on the mirror to the genus $g'$, number of boundaries $b'$ and cross-caps $c'$ on the orientifold of the mirror. We just need to note that in each case the Euler characteristic has different contributions:
\beq
\chi (\Sigma )= 2-2g \quad ; \quad  \chi (\Sigma ')= 2-2g'-b'-c' \ .
\eeq
Hence, (\ref{eq:euler}) implies
\beq
g=2g'+(b'-1)+c' \ . \label{eq:genus}
\eeq
This equation sets bounds on how the orientifold cuts the mirror surface for a given singularity.

\bigskip

\subsection{Singularities compatible with orientifold points}
\label{sec:criterion-points}

\begin{figure}[htb]
\begin{center}
\hspace*{5pt}\includegraphics[scale=.25]{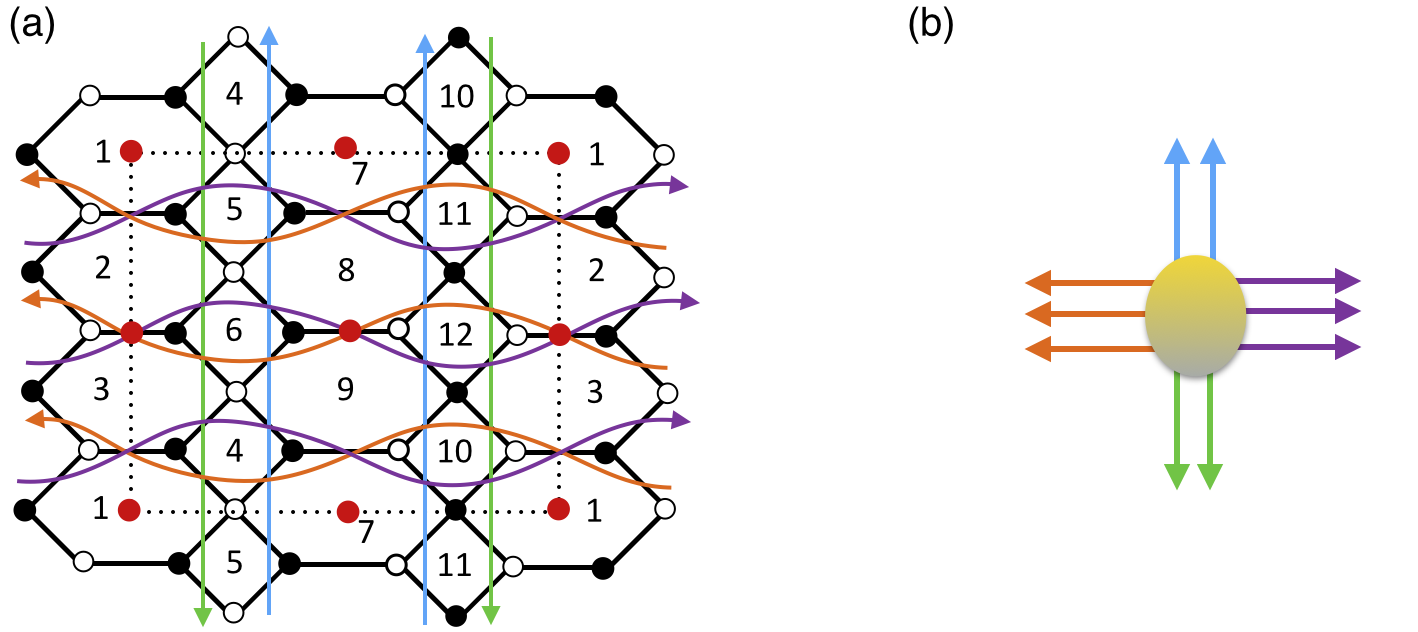} 
\caption{\small (a) Dimer for the $\mathbb{Z}_2\times \mathbb{Z}_3$ orbifold of the conifold with orientifold points in red  and its zig-zag paths. Zig-zag paths with the same winding numbers have the same colour. (b) External legs of the web diagram of the previous singularity. }
\label{fig:conifz2z3a}
\end{center}
\end{figure}

As we did for the orientifold line case, we start with an example in order to derive a general criterion  which tells what singularities accept orientifold points.  The singularity we will use for our example is a $\mathbb{Z}_2\times \mathbb{Z}_3$ orbifold of the conifold, whose dimer together with the zig-zag paths and orientifold points are shown in  figure \ref{fig:conifz2z3a}(a). The $(p,q)$'s of the singularity are shown in figure \ref{fig:conifz2z3a}(b).   

The first observation to be made is that for each field mapped to itself, we have two zig-zags that pass through the field.\footnote{Non-classical phases of dimers allow for setups where more zig-zags cross each other on top of a fixed point \cite{Garcia-Etxebarria:2015hua} due to gauge groups at strong coupling. As explained in the introduction, our focus goes on classical phases so we restrict ourselves to dimers where this statement holds.   \label{footn:strong}} Also, each zig-zag passes through two different fixed points. From here, we  see that the numbers of zig-zags and fields that are mapped to themselves is the same. Therefore, since dimers with orientifold points have four fixed points, the number of fields (and also zig-zags) mapped to themselves will be between zero and four. 
\beq
n_Z=n_F\leq 4 \label{eq:zigzag-field}
\eeq

Next, we deal with the effect of the orientifold on the zig-zags. As happened for orientifold lines, in this case we will also separate the procedure in two steps. First, we need to invert the direction of zig-zags, as in (\ref{eq:inversion}), and then we make use of the geometrical action on the dimer, which inverts both coordinates
\beq
\left\lbrace \left( \begin{matrix}
-p_i \\ -q_i \end{matrix} \right)\right\rbrace_{i=1, ..., Z}\quad \rightarrow \quad \left\lbrace \left( \begin{matrix}
p_i \\ q_i \end{matrix} \right)\right\rbrace_{i=1, ..., Z} .
\eeq
We see that the orientifold action thus leaves the $(p,q)$'s of zig-zags invariant, and thus does the same with the set of all external legs of the web diagram. 
\beq
\Omega : \  \left\lbrace \left( \begin{matrix}
p_i \\ q_i \end{matrix} \right)\right\rbrace_{i=1, ..., Z}\quad \rightarrow \quad \left\lbrace \left( \begin{matrix}
p_i \\ q_i \end{matrix} \right)\right\rbrace_{i=1, ..., Z} \ . 
\eeq
This trivial action on the set of all $(p,q)$'s does not imply that the orientifold action is in fact trivial. We can observe in figure \ref{fig:conifz2z3a} that the geometric action does relate zig-zag paths among them. These relations are of two types: either a zig-zag path crosses  fixed points and is mapped to itself, or two zig-zag paths of the same $(p,q)$ are orientifold images of one another. In the example of figure \ref{fig:conifz2z3a} we see that both zig-zag paths in blue and in green fall on the second category, whereas both for zig-zags in orange and in purple we find that one zig-zag of each color is mapped to itself while the rest are mapped by pairs. The underlying reason for this  are the limitations to arrange zig-zag paths of the same $(p,q)$ in a way compatible with orientifold points, as we now explain. 

\begin{figure}[htb]
\begin{center}
\hspace*{5pt}\includegraphics[scale=.27]{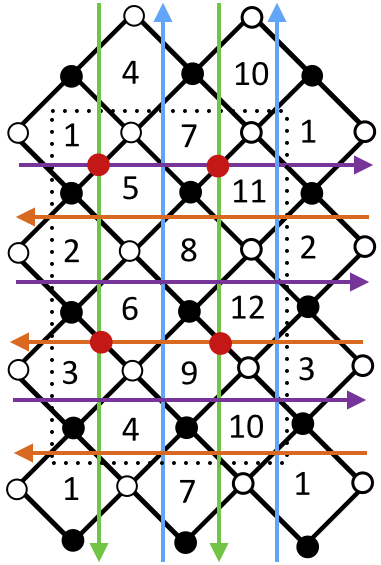} 
\caption{\small Dimer corresponding to the $\mathbb{Z}_2\times \mathbb{Z}_3$ orbifold of the conifold in a different toric phase to the one shown in figure \ref{fig:conifz2z3a}(a). We see that in this toric phase the number of zig-zags mapped to themselves is $n_Z=4$, whereas in figure \ref{fig:conifz2z3a}(a) we had $n_Z=2$. The difference between both cases is that in the previous toric phase the two zig-zags in green where orientifold images of one another, and this time each of them is mapped to itself. This is possible for this singularity because the number of zig-zags in green is even, and so accepts two types of zig-zag configurations compatible with the orientifold $\mathbb{Z}_2$ action.}
\label{fig:conifz2z3b}
\end{center}
\end{figure}

We must distinguish between two cases: on the one hand, when the number of zig-zags with same $(p,q)$ is odd, one zig-zag must necessarily get mapped to itself and the rest can be arranged in pairs, as happens for the orange and purple ones in our example. On the other hand, if the number of zig-zags with same $(p,q)$ is even two possibilities exist: one of them is arranging them in pairs as the green and blue zig-zags in figure \ref{fig:conifz2z3a}; but it is also possible to place two of them on top of fixed points and then arrange the rest in pairs. 
We illustrate this case with the dimer in figure \ref{fig:conifz2z3b}, where we have two zig-zags with the same $(p,q)$ in green that pass through orientifold points. 

Another important point can be obtained from the dimer in figure \ref{fig:conifz2z3b}. We see that it corresponds to another toric phase of the $\mathbb{Z}_2\times \mathbb{Z}_3$ orbifold of the conifold, but now we have $n_Z=4$ zig-zags falling on top of orientifold points instead of the previous $n_Z=2$. This means that it is possible to have different values of $n_Z$ for a given singularity. This general property sometimes  requires of different toric phases of a given singularity in order to obtain different values of $n_Z$, but this is not always the case, as we will show shortly.

With the information gathered from the examples, we are in a good situation to get some general properties about toric CY singularities accepting orientifold points. We just showed that for an odd number of zig-zags with the same $(p,q)$ one of them will be mapped to itself if the dimer has fixed points. This sets the minimum number of zig-zag paths (and so fields) mapped to themselves on a dimer with orientifold points. This can be made more precise by splitting the set of all zig-zags of a given singularity into subsets such that  all zig-zags in each subset have the same winding numbers $(p,q)$:
\beqa
\left\lbrace \left( \begin{matrix}
p_i \\ q_i \end{matrix} \right)\right\rbrace_{i=1, ..., Z} =\displaystyle  \left\lbrace Z_k \times\left( \begin{matrix}
p_k \\ q_k \end{matrix} \right)\right\rbrace _{k=1,..., k_S} \ .
\eeqa
Here $k_S$ is the number of subsets and $Z_k$ is the number of elements on the subset with label $k$, i.e. the number of zig-zags with winding numbers  $(p_k,q_k)$. Using this splitting of the set of all zig-zag paths, we can now talk about odd (even) subsets  when the subset has $Z_k $ odd (even). The number of odd subsets was shown above to set the minimum number of zig-zags mapped to themselves $n_Z^{min}$,  that we define as
\beqa
n_Z^{min}=\sharp (\text{Odd } Z_k ) \ .
\eeqa 
This number, together with the fact that dimers with orientifold points have four fixed points can be used to give the following rule:
\begin{flushleft}
\textbf{\underline{Criterion for toric CY singularities accepting orientifold points:}}  A toric CY singularity can have orientifold points on a classical phase of its dimer \cite{Garcia-Etxebarria:2015hua}   if $n_Z^{min}\leq 4$. For singularities with more odd subsets of zig-zags, it is not possible to have classical phases with orientifold points.   
\end{flushleft}

Furthermore, from figure \ref{fig:conifz2z3b} we see that some singularities are compatible with orientifold points and different numbers $n_Z$ of zig-zag paths mapped to themselves. This possibility depends on the value of $n_Z^{min}$ of each singularity, so we now explain what the possibilities  for each value of $n_Z^{min}$ are. We first deal with the $n_Z^{min}$ odd cases since they are more restrictive, and then study the even  $n_Z^{min} $  cases.  \begin{itemize}

\item \textbf{There are no singularities with $n_Z^{min}=1$.} The set of all $(p,q)$'s of a given singularity sums up to zero.  If $n_Z=1$ would be possible, this means that after subtracting a unique zig-zag the rest of zig-zags can be arranged in pairs, where each pair has two zig-zags with same $(p,q)$.  The sum of all these zig-zags would then have two even entries. But this sum should also be minus the $(p,q)$ of the zig-zag that is mapped to itself. Since $(p,q)$ are winding numbers, they must be mutually prime numbers, so this makes no sense.  
\item  \textbf{Singularities with $n_Z^{min}=3$   accept $n_Z=3$. }  If a singularity has odd $n_Z^{min}$ , it means that after taking $n_Z^{min}$ zig-zags the rest can be arranged by pairs of the same $(p,q)$. Therefore, in these cases $Z$ is odd. Furthermore, since the number of zig-zags $Z$ and that of gauge groups $G$ of a singularity are related by its toric diagram as observed in \cite{Hanany:2005ss},
\beq
G=Z+2g-2 \ , \label{eq:gauge-zigzag}
\eeq
we have that odd $Z$ also implies odd $G$. And odd $G$ can only be compatible with orientifold points if $n_G$, and so $n_F=n_Z$, is odd, which is our starting point.  Finally, noting that  $n_Z \leq 4$ and $n_Z\neq 1$, we find that the only singularities compatible with an odd number of zig-zags mapped to themselves are those with $n_Z=n_Z^{min}=3$. In figure \ref{fig:L232-points} we provide an example of this kind.
\begin{figure}[htb]
\begin{center}
\hspace*{5pt}\includegraphics[scale=.25]{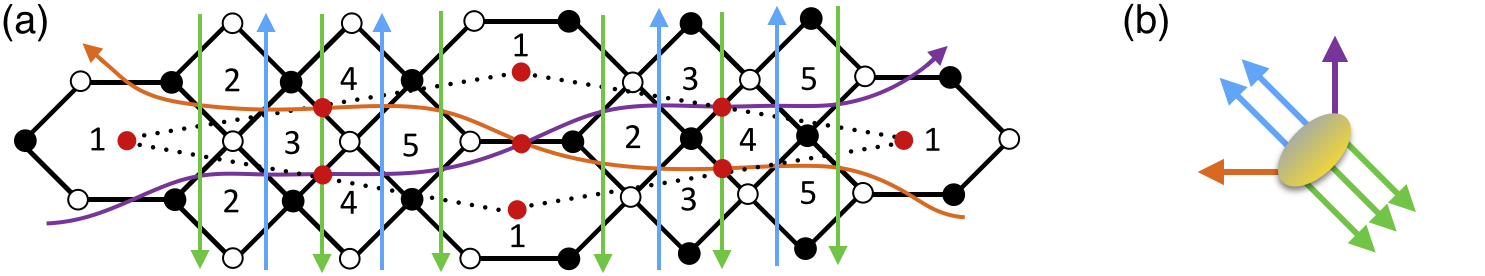} 
\caption{\small (a) Dimer of $L^{2,3,2}$ with orientifold points in red. Zig-zag paths in orange, purple and one of those in green are mapped to themselves while the rest are mapped by pairs. (b) External legs of the web diagram. It is easy to see that the subsets in orange, purple and green are the ones with an odd number of zig-zags, and thus have one element of each subset mapped to itself.}
\label{fig:L232-points}
\end{center}
\end{figure}

\item \textbf{Singularities with $n_Z^{min}=0$   accept both $n_Z=0$ and $n_Z=4$.} In this cases all subsets have an even number of elements. It is always possible  to have $n_Z=0$. No zig-zag is mapped to itself, and so no field: all fixed points fall on top of gauge groups. An example of this kind is shown in figure \ref{fig:C3z2z2}(a).
\begin{figure}[htb]
\begin{center}
\hspace*{5pt}\includegraphics[scale=.3]{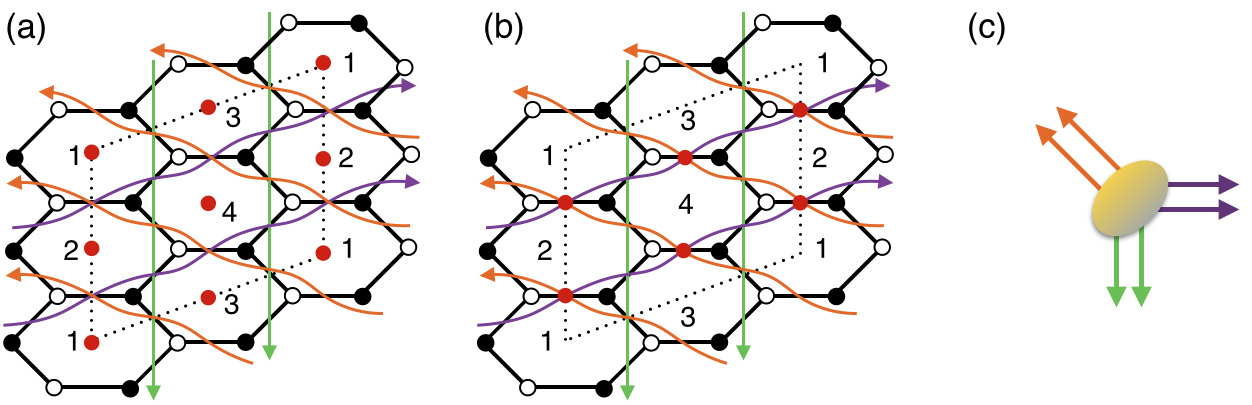} 
\caption{\small (a) Dimer of the $\mathbb{C}^3/(\mathbb{Z}_2\times\mathbb{Z}_2 )$ orbifold with orientifold points in red in a configuration where $n_Z=0$.  (b) The dimer in (a), but this time fixed points fall on top of fields so that $n_Z=4$. (c) External legs of the web diagram. Note that all zig-zags can be arranged by pairs with same $(p,q)$.}
\label{fig:C3z2z2}
\end{center}
\end{figure}

Following the ideas on the $n_Z^{min}=3$ case,  $n_Z^{min}$ even implies both $Z$ and $G$ to be even. This leaves the possibility of even values of $n_Z$ when $n_Z^{min}=0$. $n_Z=2$ turns out   not be  an option.  For this to happen  in a singularity with $n_Z^{min}=0$, the two zig-zags that would be mapped to themselves would have the same $(p,q)$, and would have to intersect each other. Zig-zags with the same $(p,q)$ never intersect, so we conclude that this is not an option. $n_Z=4$ instead is always an option, that sometimes requires the dimer to be in a different toric phase compared to the one allowing $n_Z=0$. We illustrate this with figure \ref{fig:C3z2z2}(b), where the dimer of $\mathbb{C}^3/(\mathbb{Z}_2\times\mathbb{Z}_2 )$ shows to allow both $n_Z=0$ and $n_Z=4$ in (a) and (b) respectively. As an observation, note that unlike in the case of the $\mathbb{Z}_2\times\mathbb{Z}_3$ orbifold of the conifold studied above, here both possibilities for $n_Z$ are compatible with the same toric phase. 
 
\item \textbf{When $n_Z^{min}=2$, both  $n_Z=2$ and $n_Z=4$ are possible.} This time, there is no argument preventing that both even values of $n_Z$ are possible for a singularity with $n_Z^{min}=2$. As we saw in our example, sometimes different toric phases are necessary for the different values of $n_Z$.
\item \textbf{For  singularities with $n_Z^{min}=4$ we have $n_Z=4$.} Of course, no other possibility exists in this case.
\end{itemize}

We put together all these cases on the following table:  
  \begin{center}
  \begin{tabular}{ | c | c | }
    \hline  
    $n_Z^{min}$ & Possible $n_Z=n_F$ \\ \hline\hline
    0 & 0 \& 4  \\ \hline 
        1 & -  \\ \hline 
            2 & 2 \& 4  \\ \hline 
                3 & 3  \\ \hline
                    4& 4 \\  \hline
                   $>$4 & - \\ \hline 
                      \end{tabular} 
  \end{center}

As it happened for orientifold lines, in the case of orientifold points one finds that usually  not all toric phases are compatible with orientifold points, or not all phases accept certain $n_Z$. The best way of obtaining a dimer compatible with orientifold points and $n_Z$ for a given singularity, is to use the \textit{fast inverse algorithm} \cite{Hanany:2005ss} explained in section \ref{sec:zigzag}. This time, our advise is to put first the zig-zags that are mapped to themselves and thus fall on top of orientifold points, and leave the rest for the next steps.

\bigskip

\subsubsection*{The mirror perspective}
\label{sec:criterion-points-mirror}

This time we will use the mirror perspective for different purposes. We will first  derive many of the properties we already obtained from the mirror perspective. Then, we will carry out a similar analysis to the one done for orientifold line case. Once again, these ideas are complementary to the main text and the uninterested reader may skip them. 

The constrains of the previous section can be well understood in terms of the mirror. The key feature to derive the conclusions is that two zig-zags (punctures on the mirror) with the same $(p,q)$'s will never cross  each other on the dimer. Recall that as explained in section \ref{sec:dimer-mirror},  in the tiling of the mirror surface zig-zags are faces bounding the corresponding puncture, with the boundary given by a set of edges and vertices corresponding to the fields and superpotential terms that the zig-zag path touches on the dimer. Therefore, the faces bounding two punctures  with the same $(p,q)$ on the mirror surface will never share an edge. From this fact it follows that  in any boundary of $\Sigma '$ it is necessary to have  at least two zig-zags and two fields mapped to themselves.  This fact provides a nice description in the mirror of some   constrains we gave above:\begin{itemize}
\item \textbf{$n_Z^{min}=0$ implies $n_Z\neq 2$.}  If only two punctures are mapped to themselves, the corresponding tiles in $\Sigma$ must be touching each other. Therefore, it is not possible that they have the same $(p,q)$'s. For a singularity with $n_Z^{min}=0$, to have $n_Z=2$ we would need that two punctures with same $(p,q)$'s are mapped to themselves and thus touching each other, and this is impossible as we just explained. 
\item  \textbf{$n_Z =1$ is impossible.} A zig-zag never crosses itself. Equivalently, there is no edge on the tiling of $\Sigma$ touching the same tile twice.  Therefore, it is impossible that on the mirror of the orientifold geometry there is only one puncture that falls on top of the boundary in $\Sigma '$. 
\item \textbf{$Z$ odd  only accepts $n_Z$ odd.} If the tiling of $\Sigma$ has an odd number of punctures, once we take the orientifold action the only way of arranging the punctures on a $\mathbb{Z}_2$ invariant way is by putting an odd number of them on top of the O6-plane, giving the boundary of $\Sigma '$, and then putting the rest of punctures on the \textit{bulk} of $\Sigma '$ and its orientifold image. This can also be derived from (\ref{eq:all-relations}): for $Z'$ to be an integer there is no other chance.
\item \textbf{$Z$ even  only accepts $n_Z$ even.} The argument is the same we just gave for $Z$ odd, but now with even $Z$.
\end{itemize}

The distinction about the punctures on the boundary and the \textit{bulk} of $\Sigma '$ will be  relevant when we deal with complex deformations compatible with orientifold points that we will study on section \ref{sec:deform-orientifolds}.

\medskip

\begin{figure}[htb]
\begin{center}
\hspace*{5pt}\includegraphics[scale=.23]{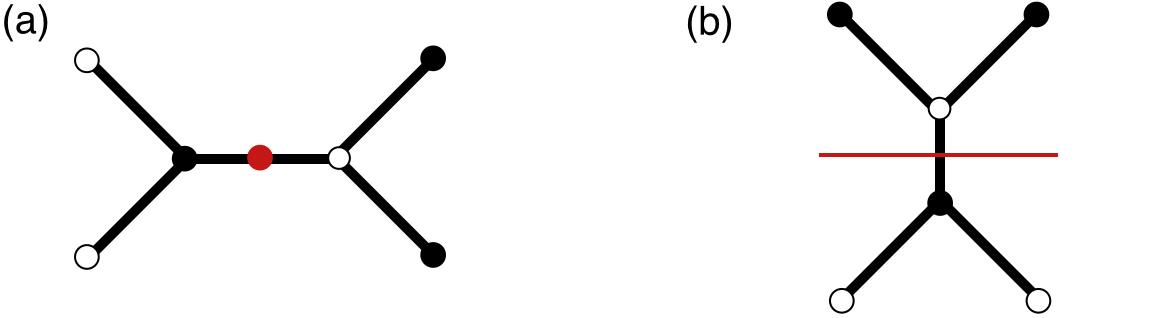} 
\caption{\small (a) Part of a dimer with orientifold points where we see a field that is mapped to itself and superpotential terms and other fields that are mapped by pairs under the orientifold action. (b) On the mirror of the previous part of the dimer, the only way to have the field mapped to itself as well as the right mapping between the vertices is to place the field perpendicular to the boundary.}
\label{fig:boundary-points}
\end{center}
\end{figure}

Now, we once again perform an analysis about   some topological properties of the orientifold of the mirror surface $\Sigma'$. Start by taking the orientifold action on the dimer, so the numbers of gauge groups, fields, etc. are reduced according to (\ref{eq:all-relations}). The difference this time comes on the relations between the numbers of elements mapped to themselves. From section \ref{sec:dimer-orientifolds} we know that no vertex is mapped to itself by orientifold points. This has consequences on the tiling of the mirror, since the only way to have a field on the boundary of $\Sigma '$ in a way such that the vertices on its sides are mapped to themselves is by placing the edge perpendicular to the boundary, as in figure \ref{fig:boundary-points}.  

This implies that there will be one field per each zig-zag on the boundary of $\Sigma '$, in agreement with the conclusion from the dimer given in (\ref{eq:zigzag-field}). Summing up
\beq
n_Z=n_F\quad \quad ; \quad \quad n_T=0 \ . \label{eq:nrelations2}
\eeq

Our next step is to proof that dimers with orientifold points live on spheres. Being more concrete, they live on spheres with four punctures (the fixed points). In order to proof this, let us consider a case where we can put all fixed points on top of faces of the dimer
\beq
G'=\dfrac{G+4}{2} \quad ; \quad  T'=\dfrac{T}{2} \quad ; \quad  F'=\dfrac{F}{2}\ .
\eeq
When we compute the Euler characteristic for this surface, fixed points fall on top of faces that we must count normally, so 
\beq
\chi (\mathbb{T}^2/\Omega_{\text{fixed points}} ) =G'+T'-F'=\dfrac{G+T-F+4}{2}=2 \ ,
\eeq
which is the Euler characteristic of the sphere. 

On the mirror of the orientifold theory there is a  boundary\footnote{In the $n_Z=0$ case there is no boundary, and in the $n_Z=2$ case there might be two boundaries. For any case,  equation (\ref{eq:pts-mirror}) holds.} that crosses the edges orthogonally, so the Euler characteristic of the orientifold of the mirror is computed in the regular way
\beq
\chi ( \Sigma ' ) =Z'+T'-F' \ . \label{eq:pts-mirror}
\eeq
This allows us to relate the initial surface $\Sigma$ and the one after taking the orientifold $\Sigma '$. We start by writing (\ref{eq:gauge-zigzag}) in terms of the parameters on the orientifold of the theory
\beq
2G'-n_G=2Z'-n_Z +2g-2 \ .
\eeq
Summing $2(T'-F')+n_F$ on both sides and using (\ref{eq:all-relations}), (\ref{eq:nrelations2}) and (\ref{eq:pts-mirror}) this is
\beq
0=  G+T-F =2 (G'+T'-F') - (n_G-n_F) = 2\chi (\Sigma ' ) -\chi (\Sigma ) \ .
\eeq
So for orientifold points we also have $2\chi (\Sigma ' ) =\chi (\Sigma )$ and so (\ref{eq:genus}) holds.

\medskip

Another very important conclusion can be made from the fact that the tile corresponding to a puncture in $\Sigma$ never touches itself is that for orientifold points there is an upper bound on the possible number of boundaries $b'$. Since each boundary involves at least two punctures, and the maximum number of punctures mapped to themselves is 4, the mirror $\Sigma '$ of a dimer with orientifold points can have a maximum of 2 boundaries. 

\bigskip

\section{Deformations of dimers with orientifolds}
\label{sec:deform-orientifolds}

Now that we have criteria to tell if a toric CY singularity is compatible with orientifold actions, and if these correspond to fixed points/lines on the dimer, we can tell  which singularities are compatible both with the orientifold action and complex deformations. As reviewed in section \ref{sec:dimer-deform},  complex deformations correspond to the removal of subwebs in equilibrium from the web diagram. The physics behind this was better understood from the mirror perspective in \cite{GarciaEtxebarria:2006aq}. Combining the above knowledge with this idea, we will now   provide criteria to tell which singularities are compatible with both orientifolds (of both kinds) and complex deformations.

\bigskip

\subsection{Deformations  compatible with orientifold lines}
\label{sec:deform-lines}

In section \ref{sec:criterion-lines} we saw that the best way of finding out if a singularity is compatible with orientifold leaving fixed lines on the dimer was to take the most symmetric representative of the web/toric diagram of the singularity (without the orientifold) and see if it is compatible with a $\mathbb{Z}_2$ action that leaves a line invariant. 

If we start with a singularity describing the UV of a warped throat compatible with orientifold lines, the IR physics after performing the complex deformation must correspond to another singularity compatible with the same $\mathbb{Z}_2$ action both on the dimer and on the web/toric diagram. Therefore, the way to deform the singularity must respect the $\mathbb{Z}_2$ symmetry. This means that the sub-web in equilibrium that we remove from the  web diagram on the UV must also be symmetric with respect to the $\mathbb{Z}_2$ action: this way the web diagram describing the IR preserves the same symmetry. 

From the mirror perspective the deformation process corresponds to removing punctures from the tiling of $\Sigma$, or $\Sigma '$ for the orientifold of the theory. For orientifold lines we find that the deformation can happen on the bulk of $\Sigma '$, on its boundary (this is the 1-cycle wrapped by the O6-plane), or in both at the same time. We illustrate these possibilities with an example of each kind. 

\begin{figure}[htb]
\begin{center}
\hspace*{5pt}\includegraphics[scale=.25]{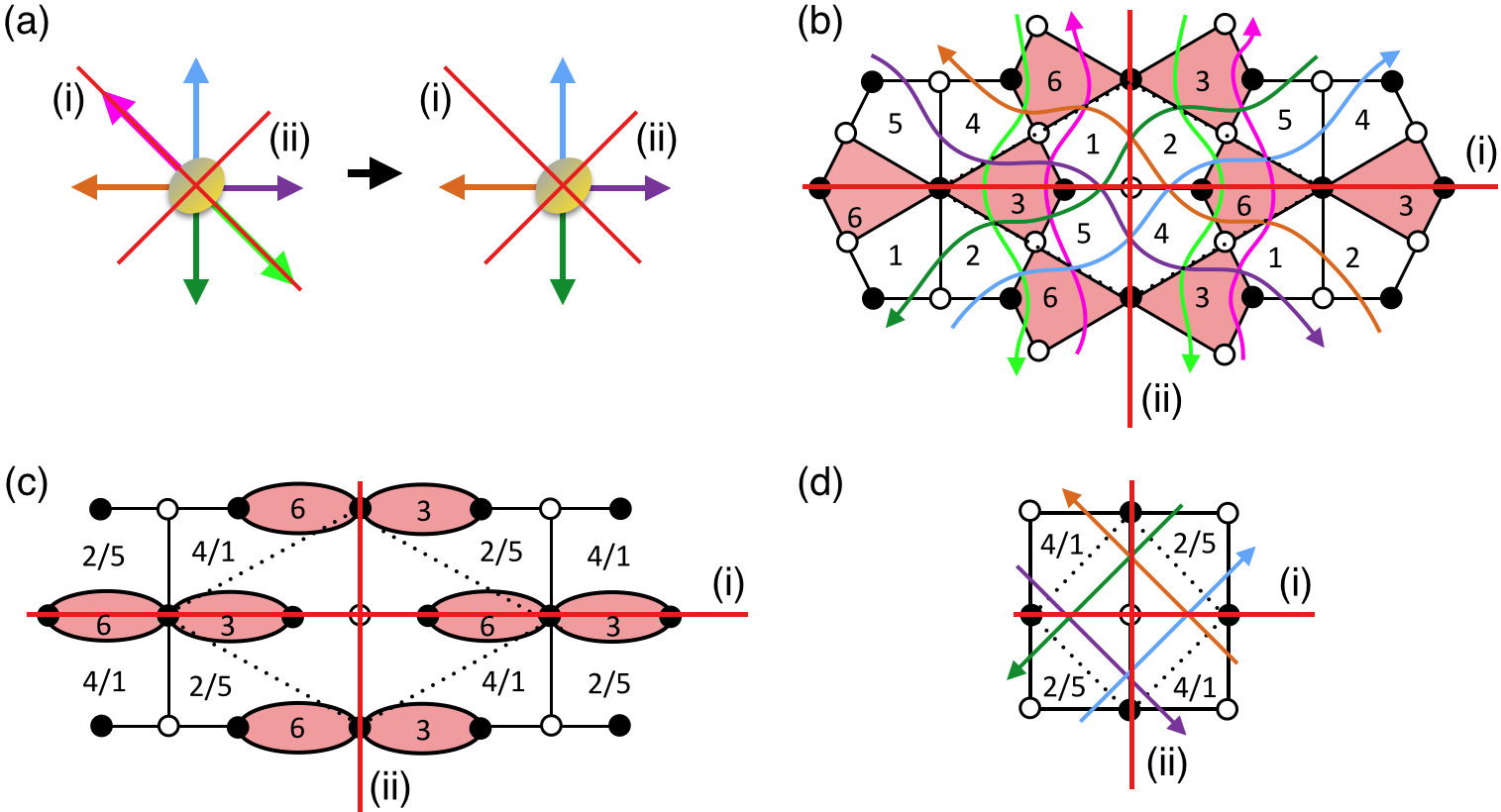} 
\caption{\small (a) External legs of the dP$_3$ theory and the conifold theory. It is easy to see that the complex deformation relating both corresponds to removing legs in light green and pink from the former. Two  orientifold lines are compatible with both singularities, we represented their action with red lines. In case (i) we see that the removed zig-zags are invariant under the orientifold action and so the corresponding punctures lie on the boundary of $\Sigma '$; whereas in case (ii) the removed zig-zags are mapped between themselves and thus they fall on the bulk of $\Sigma '$. (b) Dimer of the dP$_3$ theory with the zig-zags in different colors. The straight lines in red (i) and (ii) correspond to the two possible orientifold lines compatible we just mentioned. We are interested on the deformation involving the removal of the zig-zag paths in light green and pink, and so the confinement of gauge groups with labels 3 \& 6, that we shaded in red.  (c) Intermediate step in the confinement of gauge groups 3 \& 6, showing that it involves higgsing groups 1 \& 4 on the one hand and 2 \& 5 on the other hand to their diagonals.  (d) The dimer after the deformation process together with the two possible orientifold lines.  }
\label{fig:deform-lines-a}
\end{center}
\end{figure}

First, in figure  \ref{fig:deform-lines-a} we consider the transition from the del Pezzo 3 (dP$_3$) theory to the conifold studied in \cite{Franco:2005fd}. This transition is compatible with two different types of orientifold lines, that correspond to removing  punctures either only on the boundary of $\Sigma '$  (case (i) of the figure) or only on the bulk of this surface (case (ii) of the figure). 

For completeness, let us mention that these cases of deformations sometimes involve  strong dynamics of gauge groups of the $SO$ or/and $USp$ kind \cite{Intriligator:1995id,Intriligator:1995ne}.  For example, in the case (i) of figure \ref{fig:deform-lines-a} we find that both gauge groups 3 and 6 are of one of these types. Similar processes   were described in \cite{Franco:2015kfa}, where it could be seen that the fixed lines do not change the diagrammatic description of the deformation process.

\begin{figure}[htb]
\begin{center}
\hspace*{5pt}\includegraphics[scale=.33]{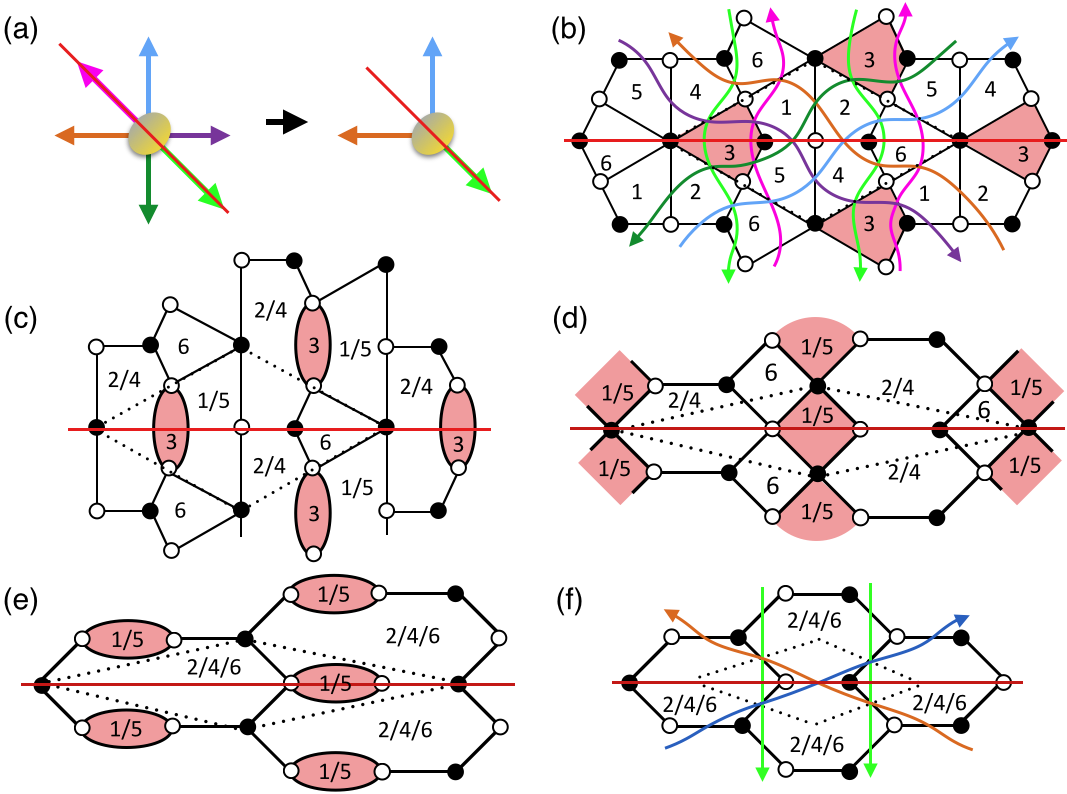} 
\caption{\small  (a) External legs of the dP$_3$ theory and the $\mathbb{C}^3$ theory. The complex deformation relating both corresponds to removing legs in dark green, purple and pink from the former. The effect of the orientifold line is shown in red, and can be seen to be compatible with the removal of the subweb.   (b) Dimer of the dP$_3$ theory with the zig-zags in different colors. We are interested on the deformation involving the removal of the zig-zag paths in dark green, purple and pink, and so the confinement of gauge groups with labels 1, 3 \& 5. Since they touch each other, this deformation was shown in \cite{Franco:2005fd} to require more than one step. We start by first confining gauge group 3,  shaded in red.  (c) Intermediate step in the confinement of gauge group 3, showing that it involves higgsing group 1  and its orientifold image 5  from being of $SU$ type to $SO/USp$ type depending on the orientifold line charge. The condensation process makes the (anti)symmetric field $X_{15}$ massive together with the (anti)symmetric meson $M_{51}$, so they get a (anti)symmetric \textit{vev} leaving a $SO/USp$  gauge factor. The same happens to gauge factor 2 and its image 4. (d) The dimer after the first confinement process. Now gauge group 1/5 confines, so we shaded it in red. (e) The next intermediate process,  where we see the higgsing of gauge groups 2/4 \& 6 to the diagonal. (f) Final dimer, corresponding to $\mathbb{C}^3$ theory with its zig-zags paths.}
\label{fig:deform-lines-b}
\end{center}
\end{figure}

Our last example of a deformation with orientifold lines mixes the two previous  cases: it involves zig-zags both in the bulk and in the boundary of $\Sigma '$. This is shown in figure \ref{fig:deform-lines-b} .  Also in this case one finds strong dynamics of gauge groups of he $SO$ or/and $USp$ kind, but this time they involve a new type of phenomenon. In the confinement of gauge group 3 we see that it higgses e.g.   group 1 (and its orientifold image 5)  from being of the $SU$ type to being of the $SO$/$USp$ type. This happens when the massive mesons of the confining group get a vev. These transform in the (anti)symmetric representation of gauge group 1 depending on the orientifold charge. The outcome is the same diagrammatic evolution of the dimer as the one without the orientifold line, that respects the $\mathbb{Z}_2$ symmetry. 

If one   intends to UV complete a singularity accepting orientifold lines on the dimer, the recipe is then pretty simple.  One must  first find the $\mathbb{Z}_2$ action leaving a fixed line on the web diagram   and then add external legs in a symmetric way with respect to this fixed line.  Finding the toric phase of the IR dimer resulting from the deformation may not be an easy task, therefore, we recommend to start from the IR theory,  add the subweb in equilibrium to the IR web diagram, and then   build the dimer of the UV theory using the  \textit{fast inverse algorithm} in \cite{Hanany:2005ss} and following our recommendations above. Finally, one needs to study the duality cascade and  the deformation on the dimer.

\bigskip

\subsection{Deformations compatible with orientifold points}
\label{sec:deform-points}

Unlike singularities compatible with orientifold lines on their dimer, those accepting orientifold points turn out to require a more careful analysis and are in a sense more restrictive with respect to complex deformations.  In section \ref{sec:criterion-points} we saw that the possibility of having orientifold points for a given toric singularity depends on the multiplicities of zig-zag paths with the same $(p,q)$ winding numbers: there will be one zig-zag path mapped to itself per each set of zig-zags with same $(p,q)$ and odd multiplicity. This sets an upper bound on the number of zig-zags mapped to themselves, that can by no means be more than four.

The starting point is to note that complex deformations are described as condensation of certain gauge groups on the dimer. These gauge groups need to be rectangles in order to have at certain point $N_c\geq N_f$. As explained in section \ref{sec:dimer-orientifolds} rectangles do not accept orientifold points on top of them.  Therefore, when orientifold points are present for every gauge group condensing there will be another one (its orientifold image) doing the same. At this point one can think of two possibilities: the groups we want to confine could be touching each other and thus touching a fixed point, or they could be away from the orientifold point and thus not touching each other. These correspond to D6-branes on the mirror surface $\Sigma'$, that do touch the boundary in the first case, but do not touch it on the second one. This means that the first case translates to a deformation on the boundary of $\Sigma'$, whereas the second case corresponds to a deformation on the bulk of $\Sigma'$ and its orientifold image, or equivalently, in the first case one wants to remove punctures/zig-zags that are mapped to themselves and in the second case pairs of punctures/zig-zags away from the orientifold.

\begin{figure}[htb]
\begin{center}
\hspace*{5pt}\includegraphics[scale=.27]{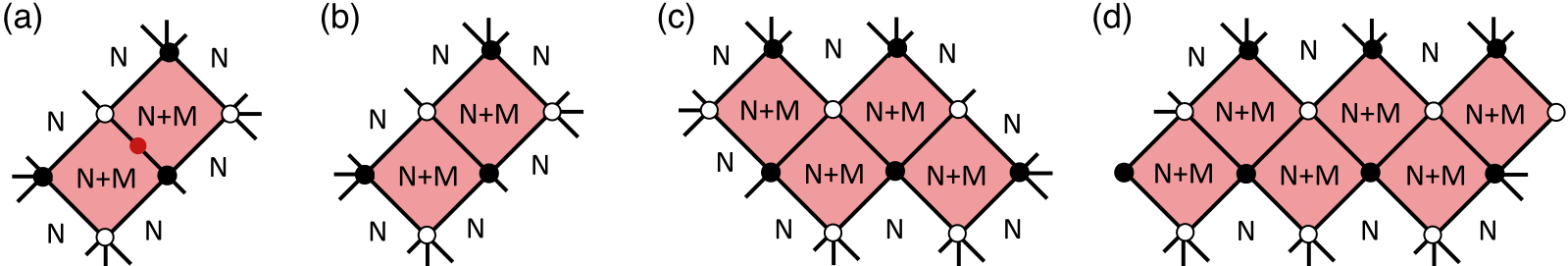} 
\caption{\small (a) In order to have a complex deformation removing punctures on the boundary of $\Sigma'$ we need gauge groups whose D6-branes on the mirror touch the orientifold. They would give a configuration like this one on the dimer, with $N$ regular branes and $M$ fractional branes on the confining groups. (b) In order to simplify the analysis, we can remove the orientifold point while keeping the $\mathbb{Z}_2$ symmetry. (c) The setup in (b) is anomalous and requires putting fractional branes in more gauge factors. (d) Following the same philosophy, the gauge factors with fractional branes keeps growing.}
\label{fig:anomalias}
\end{center}
\end{figure}

Let us first deal with the case where the confining groups do touch a fixed point  as well as its orientifold image on the dimer. We anticipate that this process is not possible. Since we have a D6-brane touching the boundary and confining, the dimer looks locally as shown in figure \ref{fig:anomalias}(a).  

\begin{figure}[htb]
\begin{center}
\hspace*{5pt}\includegraphics[scale=.16]{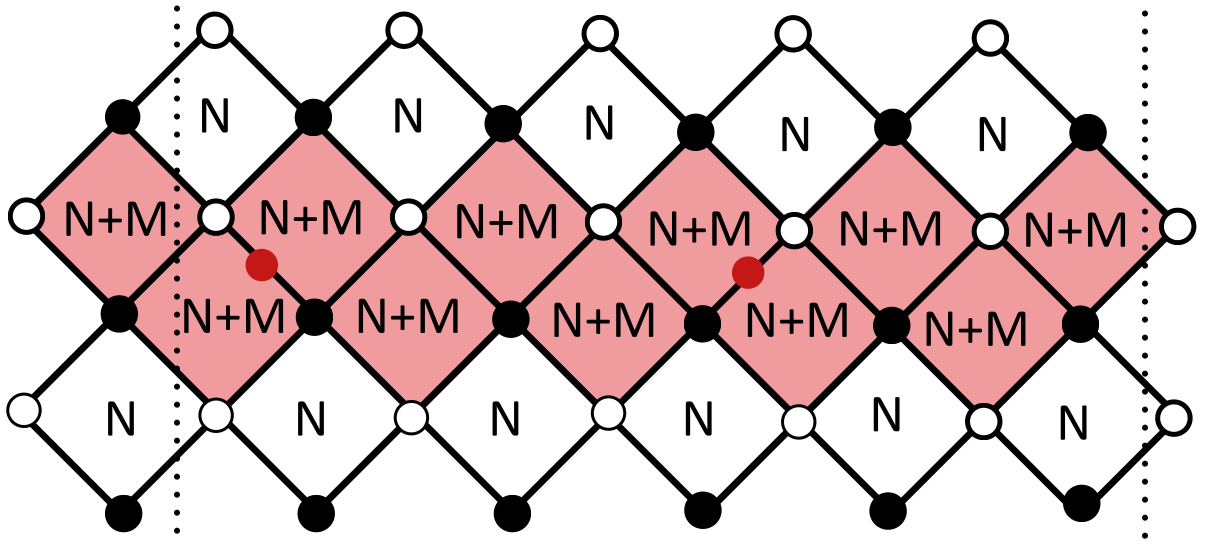} 
\caption{\small Part of a dimer describing the prototypical situation one finds when adding fractional branes until anomaly cancellation conditions are satisfied for all gauge groups on the dimer. Shaded faces are those with fractional branes, whereas the ones in white only have regular branes. For simplicity, the $\mathcal{O}(1)$ effects due to the orientifold RR charge where neglected. These type of fractional brane distribution corresponds to the so-called $\mathcal{N}=2$ fractional branes. This figure could describe part of e.g. a $\mathbb{Z}_5 \times \mathbb{Z}_n$ orbifold of the conifold.}
\label{fig:columna}
\end{center}
\end{figure}

Our argument against deformations in this case relies on the distribution of fractional branes (distribution of ranks of gauge groups) on the dimer. The orientifold has RR charge and thus contributes to the anomaly cancellation conditions, but its effects are  $\mathcal{O}(1)$  compared to those of the $N$ regular branes and $M$ fractional branes that we consider. For simplicity, it is thus better to  consider a dimer without the orientifold point contribution but preserving the $\mathbb{Z}_2$ symmetry, as in figure \ref{fig:anomalias}(b).  We will eventually find out that a deformation of the kind we would like to have is not possible in this   case, which extends to the case where the orientifold point is included. As a  starting point, consider the dimer with no fractional branes. Now, since we want to confine  two gauge groups (that are orientifold images of one another in the case with the orientifold) we put fractional branes on them, leaving the local setup in the dimer  in  figure \ref{fig:anomalias}(b). In order for these groups to be anomaly-free, it is necessary to include fractional branes in other faces of the dimer. We include these fractional branes in adjacent faces while keeping the $\mathbb{Z}_2$ symmetry, see figure \ref{fig:anomalias}(c). At this point, the faces where we just put fractional branes are anomalous, and require putting fractional branes on other gauge groups. This process follows, growing the number of gauge groups with fractional branes until one reaches the point where the bi-periodicity of the dimer makes faces with fractional branes meet other faces with fractional branes and no more steps need to be taken. This means that in the end, in the resulting anomaly-free $\mathbb{Z}_2$ configuration, one can wind at least   one of the 1-cycles of the $\mathbb{T}^2$ of the dimer. An example of this kind is shown in figure \ref{fig:columna}. These type of fractional branes were dubbed in \cite{Franco:2005zu} $\mathcal{N}=2$ branes, and it was shown that they do not lead to complex deformations. A more extreme possibility would be that the final configuration requires putting fractional branes on all faces of the dimer, which are just regular branes, and thus cannot lead to confinement. Let us emphasize that this conclusions come just from demanding that the fractional branes fall on top of adjacent gauge groups and are distributed in a $\mathbb{Z}_2$ symmetric way. From here we conclude that it is not possible to perform a complex deformation removing zig-zags that are mapped to themselves for a singularity with orientifold points.   

\begin{figure}[htb]
\begin{center}
\hspace*{5pt}\includegraphics[scale=.33]{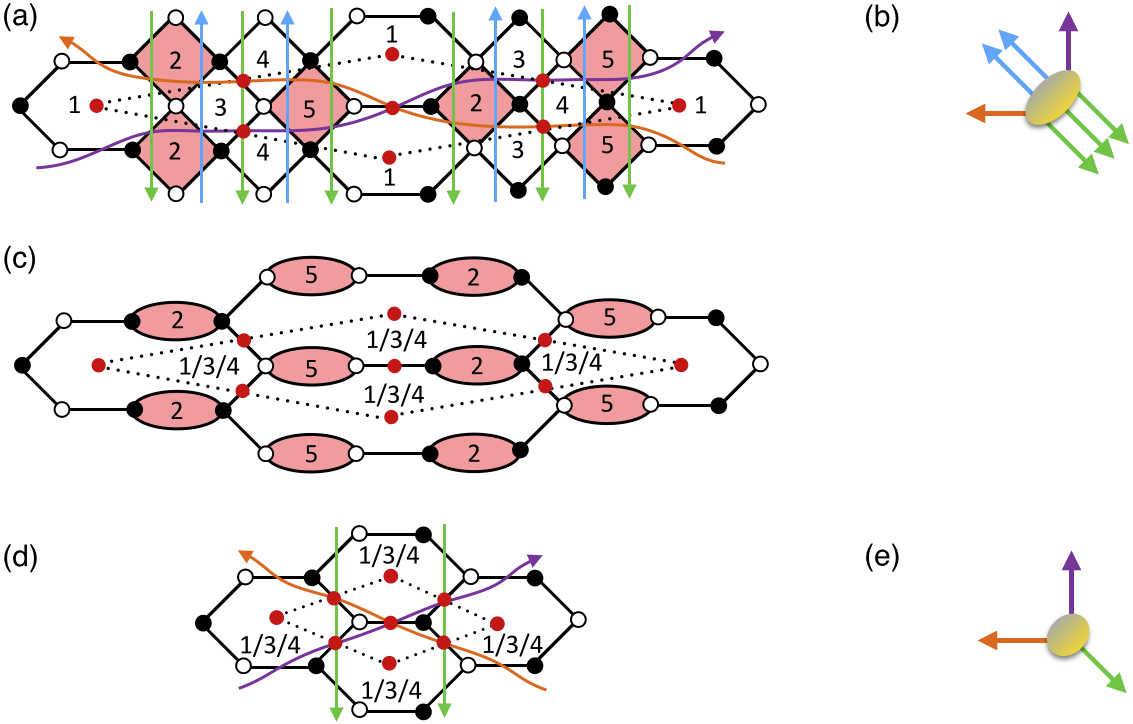} 
\caption{\small (a) Dimer diagram of $L^{2,3,2}$ with orientifold points. We want to show a deformation process to the IR by removing zig-zags on the bulk and their orientifold images, this corresponds to studying the strong dynamics of gauge group 2 and its orientifold image 5, each bound by a zig-zag in green and one in blue. We shaded these groups in red. (b) External legs of $L^{2,3,2}$. We see that one zig-zag in green plus the ones in purple and orange must be mapped to themselves ($n_Z=3$), in agreement with the dimer. Since we want to remove external legs mapped between themselves, this corresponds to the other two external legs in green plus the two in blue. Note that without the orientifold this process could be separated into two individual deformations. (c) Intermediate step in the deformation process, when groups 2 and 5 confine their mesons acquire vevs that  higgs gauge group $SU(n_1)\times USp(n_3)\times SU(n_5)$ to $USp(n_1/2=n_3=n_5/2) $ for e.g. negative orientifold charge on the fixed point on the top. (d) Dimer of the $\mathbb{C}^3$ theory with orientifold points and zig-zags after the complex deformation. (e) External legs of the $\mathbb{C}^3$ theory. By comparing it with (b) we see that the external legs missing were orientifold images of one another.}
\label{fig:deform-points-a}
\end{center}
\end{figure}

The other option is that the confining gauge groups and their images do not touch any fixed point. In this case,  the corresponding D6-branes in the mirror dual do not touch the orientifold, and so are wrapping a series of punctures in $\Sigma '$ that are on the bulk. Therefore, the  external legs removed on the deformation will  be on the bulk of $\Sigma '$ (and its orientifold image). This corresponds to removing a subweb in equilibrium that contains no zig-zag path that is mapped to itself under the orientifold action, i.e. the deformation requires removing two copies of a subweb in equilibrium from the web diagram of the UV theory. In figure \ref{fig:deform-points-a} we show an example of this kind.

A final remark goes on how to UV complete singularities with orientifold points. As done for orientifold lines, the best strategy seems to be to start with the web diagram of the IR singularity and add to it a subweb in equilibrium twice. This gives the UV singularity, whose dimer can be constructed using the \textit{fast forward algorithm } \cite{Hanany:2005ss}. The warped throat behaviour and following deformation can then be addressed by finding the RG cascade on the UV dimer and then the confinement process that gives the IR singularity.

\bigskip

\section{Conclusions}
\label{sec:conclusions}

The flexibility of warped throats as local configurations able to generate hierarchies in string compactifications motivates the systematic exploration of possible throat geometries and their properties. Recent applications exploit properties which require the presence of orientifold planes, either to remove physical degrees of freedom (by the orientifold projection) or to trigger the appearance of non-perturbative effects (due to the orientifold projection on fermion zero modes of the underlying instantons). This motivates the development of new tools to characterize orientifolded warped throats, a task we have addressed in the present paper.

Our results provide the basic criteria that determine which toric singularities admit orientifold actions of given type (with fixed lines or points in the corresponding dimer diagram), and to embed them as IR configurations of complex deformed geometries supporting orientifolded throats. The basic criteria are determined by simple properties of zig-zag paths in the dimer diagram, which cleverly encode the combinatorics of the behaviour of the singularity under the orientifold action.

Our construction reproduces the  orientifolded warped throats used in the gauge theory description of certain D-brane instantons \cite{Aharony:2007pr,Amariti:2008xu,Franco:2015kfa}, and de Sitter uplifts using string embeddings of the nilpotent goldstino \cite{Kallosh:2015nia} and DSB sectors \cite{Retolaza:2015nvh}. We hope that our developments will help in the construction of new orientifolded throats yielding similarly interesting future applications.

\bigskip

\section*{Acknowledgments}

We would like to thank I\~naki Garc\'ia-Etxebarria, Luis Ib\'a\~{n}ez and Gianluca Zoccarato 
 for useful discussions and  also Ben Heidenreich for   comments  about a previous version. We are also indebted to Sebasti\'an Franco for useful comments while developing previous projects. A. R. and A. U. are partially supported by the grants FPA2012-32828 from the MINECO, the ERC Advanced Grant SPLE under contract ERC-2012-ADG-20120216-320421 and the grant SEV-2012-0249 of the ``Centro de Excelencia Severo Ochoa" Programme.

\newpage

\bibliographystyle{JHEP}
\bibliography{mybib}

\providecommand{\href}[2]{#2}\begingroup\raggedright\begin{thebibliography}{10}

\bibitem{Kachru:2003aw}
S.~Kachru, R.~Kallosh, A.~D. Linde, and S.~P. Trivedi, {\it {De Sitter vacua in
  string theory}},  {\em Phys. Rev.} {\bf D68} (2003) 046005,
  [\href{http://arxiv.org/abs/hep-th/0301240}{{\tt hep-th/0301240}}].

\bibitem{Kachru:2003sx}
S.~Kachru, R.~Kallosh, A.~D. Linde, J.~M. Maldacena, L.~P. McAllister, and
  S.~P. Trivedi, {\it {Towards inflation in string theory}},  {\em JCAP} {\bf
  0310} (2003) 013, [\href{http://arxiv.org/abs/hep-th/0308055}{{\tt
  hep-th/0308055}}].

\bibitem{Franco:2014hsa}
S.~Franco, D.~Galloni, A.~Retolaza, and A.~Uranga, {\it {On axion monodromy
  inflation in warped throats}},  {\em JHEP} {\bf 02} (2015) 086,
  [\href{http://arxiv.org/abs/1405.7044}{{\tt arXiv:1405.7044}}].

\bibitem{Retolaza:2015sta}
A.~Retolaza, A.~M. Uranga, and A.~Westphal, {\it {Bifid Throats for Axion
  Monodromy Inflation}},  {\em JHEP} {\bf 07} (2015) 099,
  [\href{http://arxiv.org/abs/1504.02103}{{\tt arXiv:1504.02103}}].

\bibitem{Cascales:2003wn}
J.~F.~G. Cascales, M.~P. Garcia~del Moral, F.~Quevedo, and A.~M. Uranga, {\it
  {Realistic D-brane models on warped throats: Fluxes, hierarchies and moduli
  stabilization}},  {\em JHEP} {\bf 02} (2004) 031,
  [\href{http://arxiv.org/abs/hep-th/0312051}{{\tt hep-th/0312051}}].

\bibitem{Cascales:2005rj}
J.~F. Cascales, F.~Saad, and A.~M. Uranga, {\it {Holographic dual of the
  standard model on the throat}},  {\em JHEP} {\bf 0511} (2005) 047,
  [\href{http://arxiv.org/abs/hep-th/0503079}{{\tt hep-th/0503079}}].

\bibitem{Franco:2008jc}
S.~Franco, D.~Rodriguez-Gomez, and H.~Verlinde, {\it {N-ification of Forces: A
  Holographic Perspective on D-brane Model Building}},  {\em JHEP} {\bf 0906}
  (2009) 030, [\href{http://arxiv.org/abs/0804.1125}{{\tt arXiv:0804.1125}}].

\bibitem{Klebanov:2000hb}
I.~R. Klebanov and M.~J. Strassler, {\it {Supergravity and a confining gauge
  theory: Duality cascades and chi SB resolution of naked singularities}},
  {\em JHEP} {\bf 0008} (2000) 052,
  [\href{http://arxiv.org/abs/hep-th/0007191}{{\tt hep-th/0007191}}].

\bibitem{Franco:2004jz}
S.~Franco, Y.-H. He, C.~Herzog, and J.~Walcher, {\it {Chaotic duality in string
  theory}},  {\em Phys.Rev.} {\bf D70} (2004) 046006,
  [\href{http://arxiv.org/abs/hep-th/0402120}{{\tt hep-th/0402120}}].

\bibitem{Franco:2005fd}
S.~Franco, A.~Hanany, and A.~M. Uranga, {\it {Multi-flux warped throats and
  cascading gauge theories}},  {\em JHEP} {\bf 0509} (2005) 028,
  [\href{http://arxiv.org/abs/hep-th/0502113}{{\tt hep-th/0502113}}].

\bibitem{Giddings:2001yu}
S.~B. Giddings, S.~Kachru, and J.~Polchinski, {\it {Hierarchies from fluxes in
  string compactifications}},  {\em Phys. Rev.} {\bf D66} (2002) 106006,
  [\href{http://arxiv.org/abs/hep-th/0105097}{{\tt hep-th/0105097}}].

\bibitem{Aharony:2007pr}
O.~Aharony and S.~Kachru, {\it {Stringy Instantons and Cascading Quivers}},
  {\em JHEP} {\bf 0709} (2007) 060, [\href{http://arxiv.org/abs/0707.3126}{{\tt
  arXiv:0707.3126}}].

\bibitem{Amariti:2008xu}
A.~Amariti, L.~Girardello, and A.~Mariotti, {\it {Stringy Instantons as Strong
  Dynamics}},  {\em JHEP} {\bf 0811} (2008) 041,
  [\href{http://arxiv.org/abs/0809.3432}{{\tt arXiv:0809.3432}}].

\bibitem{Franco:2015kfa}
S.~Franco, A.~Retolaza, and A.~Uranga, {\it {D-brane Instantons as Gauge
  Instantons in Orientifolds of Chiral Quiver Theories}},  {\em JHEP} {\bf 11}
  (2015) 165, [\href{http://arxiv.org/abs/1507.05330}{{\tt arXiv:1507.05330}}].

\bibitem{Kallosh:2015nia}
R.~Kallosh, F.~Quevedo, and A.~M. Uranga, {\it {String Theory Realizations of
  the Nilpotent Goldstino}},  \href{http://arxiv.org/abs/1507.07556}{{\tt
  arXiv:1507.07556}}.

\bibitem{Retolaza:2015nvh}
A.~Retolaza and A.~Uranga, {\it {De Sitter Uplift with Dynamical Susy
  Breaking}},  {\em JHEP} {\bf 04} (2016) 137,
  [\href{http://arxiv.org/abs/1512.06363}{{\tt arXiv:1512.06363}}].

\bibitem{Franco:2007ii}
S.~Franco, A.~Hanany, D.~Krefl, J.~Park, A.~M. Uranga, and D.~Vegh, {\it
  {Dimers and orientifolds}},  {\em JHEP} {\bf 0709} (2007) 075,
  [\href{http://arxiv.org/abs/0707.0298}{{\tt arXiv:0707.0298}}].

\bibitem{Franco:2005rj}
S.~Franco, A.~Hanany, K.~D. Kennaway, D.~Vegh, and B.~Wecht, {\it {Brane dimers
  and quiver gauge theories}},  {\em JHEP} {\bf 0601} (2006) 096,
  [\href{http://arxiv.org/abs/hep-th/0504110}{{\tt hep-th/0504110}}].

\bibitem{Franco:2005sm}
S.~Franco, A.~Hanany, D.~Martelli, J.~Sparks, D.~Vegh, et~al., {\it {Gauge
  theories from toric geometry and brane tilings}},  {\em JHEP} {\bf 0601}
  (2006) 128, [\href{http://arxiv.org/abs/hep-th/0505211}{{\tt
  hep-th/0505211}}].

\bibitem{Kennaway:2007tq}
K.~D. Kennaway, {\it {Brane Tilings}},  {\em Int. J. Mod. Phys.} {\bf A22}
  (2007) 2977--3038, [\href{http://arxiv.org/abs/0706.1660}{{\tt
  arXiv:0706.1660}}].

\bibitem{Yamazaki:2008bt}
M.~Yamazaki, {\it {Brane Tilings and Their Applications}},  {\em Fortsch.
  Phys.} {\bf 56} (2008) 555--686, [\href{http://arxiv.org/abs/0803.4474}{{\tt
  arXiv:0803.4474}}].

\bibitem{Hanany:2005ss}
A.~Hanany and D.~Vegh, {\it {Quivers, tilings, branes and rhombi}},  {\em JHEP}
  {\bf 10} (2007) 029, [\href{http://arxiv.org/abs/hep-th/0511063}{{\tt
  hep-th/0511063}}].

\bibitem{Feng:2005gw}
B.~Feng, Y.-H. He, K.~D. Kennaway, and C.~Vafa, {\it {Dimer models from mirror
  symmetry and quivering amoebae}},  {\em Adv. Theor. Math. Phys.} {\bf 12}
  (2008), no.~3 489--545, [\href{http://arxiv.org/abs/hep-th/0511287}{{\tt
  hep-th/0511287}}].

\bibitem{Aharony:1997ju}
O.~Aharony and A.~Hanany, {\it {Branes, superpotentials and superconformal
  fixed points}},  {\em Nucl. Phys.} {\bf B504} (1997) 239--271,
  [\href{http://arxiv.org/abs/hep-th/9704170}{{\tt hep-th/9704170}}].

\bibitem{Aharony:1997bh}
O.~Aharony, A.~Hanany, and B.~Kol, {\it {Webs of (p,q) five-branes,
  five-dimensional field theories and grid diagrams}},  {\em JHEP} {\bf 01}
  (1998) 002, [\href{http://arxiv.org/abs/hep-th/9710116}{{\tt
  hep-th/9710116}}].

\bibitem{Leung:1997tw}
N.~C. Leung and C.~Vafa, {\it {Branes and toric geometry}},  {\em Adv. Theor.
  Math. Phys.} {\bf 2} (1998) 91--118,
  [\href{http://arxiv.org/abs/hep-th/9711013}{{\tt hep-th/9711013}}].

\bibitem{Franco:2005zu}
S.~Franco, A.~Hanany, F.~Saad, and A.~M. Uranga, {\it {Fractional branes and
  dynamical supersymmetry breaking}},  {\em JHEP} {\bf 0601} (2006) 011,
  [\href{http://arxiv.org/abs/hep-th/0505040}{{\tt hep-th/0505040}}].

\bibitem{GarciaEtxebarria:2012qx}
I.~Garcia-Etxebarria, B.~Heidenreich, and T.~Wrase, {\it {New N=1 dualities
  from orientifold transitions. Part I. Field Theory}},  {\em JHEP} {\bf 10}
  (2013) 007, [\href{http://arxiv.org/abs/1210.7799}{{\tt arXiv:1210.7799}}].

\bibitem{Garcia-Etxebarria:2013tba}
I.~García-Etxebarria, B.~Heidenreich, and T.~Wrase, {\it {New N=1 dualities
  from orientifold transitions - Part II: String Theory}},  {\em JHEP} {\bf 10}
  (2013) 006, [\href{http://arxiv.org/abs/1307.1701}{{\tt arXiv:1307.1701}}].

\bibitem{Garcia-Etxebarria:2015hua}
I.~García-Etxebarria and B.~Heidenreich, {\it {Strongly coupled phases of
  $\mathcal{N}=1$ S-duality}},  {\em JHEP} {\bf 09} (2015) 032,
  [\href{http://arxiv.org/abs/1506.03090}{{\tt arXiv:1506.03090}}].

\bibitem{Garcia-Etxebarria:2015lif}
I.~García-Etxebarria, F.~Quevedo, and R.~Valandro, {\it {Global String
  Embeddings for the Nilpotent Goldstino}},  {\em JHEP} {\bf 02} (2016) 148,
  [\href{http://arxiv.org/abs/1512.06926}{{\tt arXiv:1512.06926}}].

\bibitem{Hanany:2005ve}
A.~Hanany and K.~D. Kennaway, {\it {Dimer models and toric diagrams}},
  \href{http://arxiv.org/abs/hep-th/0503149}{{\tt hep-th/0503149}}.

\bibitem{Seiberg:1994pq}
N.~Seiberg, {\it {Electric - magnetic duality in supersymmetric nonAbelian
  gauge theories}},  {\em Nucl. Phys.} {\bf B435} (1995) 129--146,
  [\href{http://arxiv.org/abs/hep-th/9411149}{{\tt hep-th/9411149}}].

\bibitem{Beasley:2001zp}
C.~E. Beasley and M.~R. Plesser, {\it {Toric duality is Seiberg duality}},
  {\em JHEP} {\bf 12} (2001) 001,
  [\href{http://arxiv.org/abs/hep-th/0109053}{{\tt hep-th/0109053}}].

\bibitem{Feng:2001bn}
B.~Feng, A.~Hanany, Y.-H. He, and A.~M. Uranga, {\it {Toric duality as Seiberg
  duality and brane diamonds}},  {\em JHEP} {\bf 12} (2001) 035,
  [\href{http://arxiv.org/abs/hep-th/0109063}{{\tt hep-th/0109063}}].

\bibitem{Klebanov:2000nc}
I.~R. Klebanov and A.~A. Tseytlin, {\it {Gravity duals of supersymmetric SU(N)
  x SU(N+M) gauge theories}},  {\em Nucl. Phys.} {\bf B578} (2000) 123--138,
  [\href{http://arxiv.org/abs/hep-th/0002159}{{\tt hep-th/0002159}}].

\bibitem{Bouchard:2007ik}
V.~Bouchard, {\it {Lectures on complex geometry, Calabi-Yau manifolds and toric
  geometry}},  \href{http://arxiv.org/abs/hep-th/0702063}{{\tt
  hep-th/0702063}}.

\bibitem{Aldazabal:2000cn}
G.~Aldazabal, S.~Franco, L.~E. Ibanez, R.~Rabadan, and A.~M. Uranga, {\it
  {Intersecting brane worlds}},  {\em JHEP} {\bf 02} (2001) 047,
  [\href{http://arxiv.org/abs/hep-ph/0011132}{{\tt hep-ph/0011132}}].

\bibitem{Aldazabal:2000dg}
G.~Aldazabal, S.~Franco, L.~E. Ibanez, R.~Rabadan, and A.~M. Uranga, {\it {D =
  4 chiral string compactifications from intersecting branes}},  {\em J. Math.
  Phys.} {\bf 42} (2001) 3103--3126,
  [\href{http://arxiv.org/abs/hep-th/0011073}{{\tt hep-th/0011073}}].

\bibitem{GarciaEtxebarria:2006aq}
I.~Garcia-Etxebarria, F.~Saad, and A.~M. Uranga, {\it {Quiver gauge theories at
  resolved and deformed singularities using dimers}},  {\em JHEP} {\bf 0606}
  (2006) 055, [\href{http://arxiv.org/abs/hep-th/0603108}{{\tt
  hep-th/0603108}}].

\bibitem{Benvenuti:2005ja}
S.~Benvenuti and M.~Kruczenski, {\it {From Sasaki-Einstein spaces to quivers
  via BPS geodesics: L**p,q|r}},  {\em JHEP} {\bf 04} (2006) 033,
  [\href{http://arxiv.org/abs/hep-th/0505206}{{\tt hep-th/0505206}}].

\bibitem{Butti:2005sw}
A.~Butti, D.~Forcella, and A.~Zaffaroni, {\it {The Dual superconformal theory
  for L**pqr manifolds}},  {\em JHEP} {\bf 09} (2005) 018,
  [\href{http://arxiv.org/abs/hep-th/0505220}{{\tt hep-th/0505220}}].

\bibitem{Intriligator:1995id}
K.~A. Intriligator and N.~Seiberg, {\it {Duality, monopoles, dyons, confinement
  and oblique confinement in supersymmetric SO(N(c)) gauge theories}},  {\em
  Nucl. Phys.} {\bf B444} (1995) 125--160,
  [\href{http://arxiv.org/abs/hep-th/9503179}{{\tt hep-th/9503179}}].

\bibitem{Intriligator:1995ne}
K.~A. Intriligator and P.~Pouliot, {\it {Exact superpotentials, quantum vacua
  and duality in supersymmetric SP(N(c)) gauge theories}},  {\em Phys.Lett.}
  {\bf B353} (1995) 471--476, [\href{http://arxiv.org/abs/hep-th/9505006}{{\tt
  hep-th/9505006}}].

\end{thebibliography}\endgroup

\end{document}